\documentclass[aps, prd, twocolumn, superscriptaddress, nofootinbib]{revtex4-1}

\usepackage{bm,amssymb,slashed,graphicx,multirow,soul,mathtools,xspace,array}

\usepackage{hyperref}

\usepackage{breakurl}
\usepackage{array}
\usepackage{slashed}
\usepackage{xspace}
\usepackage{booktabs}
\usepackage{hhline}
\usepackage{multirow}
\usepackage{header3}
\usepackage{placeins}
\usepackage{physics}
\usepackage{listings}
\usepackage{gensymb}
\usepackage{lineno}
\usepackage{todonotes}

\definecolor{mygray}{rgb}{0.4,0.4,0.4}
\definecolor{mygreen}{rgb}{0,0.8,0.6}
\definecolor{myorange}{rgb}{1.0,0.4,0}

\usepackage{xpunctuate}

\newcommand{\eg}{e.g\xperiod}

\newcommand{\Bbar}{\,\overline{\!B}}
\def\B0bar{\Bbar{}^0}

\newcommand*\rhoomega{\ensuremath{{\rho\text{-}\omega}}\xspace}

\def\spnt{1}
\if\spnt1

\else

\fi

\newcommand{\GeV}{\text{GeV}}
\newcommand{\MeV}{\text{MeV}}

\newcommand{\beq}{\begin{equation}}
\newcommand{\eeq}{\end{equation}}
\newcommand{\beqa}{\begin{eqnarray}}
\newcommand{\eeqa}{\end{eqnarray}}

\DeclareSIUnit{\clight}{\text{\ensuremath{c}}}
\DeclareSIUnit[per-mode=symbol]\GeVcc{\GeV\per\clight\squared}
\DeclareSIUnit[per-mode=symbol]\MeVcc{\MeV\per\clight\squared}

\newcommand{\babar}{\mbox{\ensuremath{{\displaystyle B}\!{\scriptstyle A}{\displaystyle B}\!{\scriptstyle AR}}}\xspace}

\makeatletter
\g@addto@macro\bfseries{\boldmath}
\makeatother

\allowdisplaybreaks

\newcommand{\BFrho}{\ensuremath{\mathcal{B}(B \to \rho^0 \ell \bar \nu_\ell)}\xspace}
\newcommand{\BFrhoFit}{\mbox{\ensuremath{ \mathcal{B}(B \to \rho^0 \ell \bar \nu_\ell) = \left(1.41_{-0.38}^{+0.49}  \right) \times 10^{-4} }} }

\newcommand{\BFrhoFitValue}{\mbox{\ensuremath{ \left(1.41_{-0.38}^{+0.49} \right) \times 10^{-4} }}}

\newcommand{\BFpipiFit}{\mbox{\ensuremath{ \Delta \mathcal{B}(B \to [\pi^+ \pi^-]_S \, \ell \bar \nu_\ell)  <  0.51 \times 10^{-4} \,\, \mathrm{at} \, \, 90\% \, \mathrm{CL} }}}

\newcommand{\BFpipiFitValue}{\mbox{\ensuremath{  0.51 \times 10^{-4} \,\, \mathrm{at} \, \, 90\% \, \mathrm{CL} }}}

\newcommand{\PhaseomrhoFit}{\mbox{\ensuremath{ \phi_{\rhoomega}  = \left( -46_{\,-67}^{+155}  \right)\degree }}}

\newcommand{\PhaseomrhoFitValue}{\mbox{\ensuremath{ \left( -46_{\,-67}^{+155}  \right)\degree }}}

\begin{document}

\title{  Role of \rhoomega interference in semileptonic $B \to \pi^+ \pi^- \ell \bar \nu_\ell$ decays }

\author{Florian U.\ Bernlochner}
\affiliation{Physikalisches Institut der Rheinischen Friedrich-Wilhelms-Universit\"at Bonn, 53115 Bonn, Germany}

\author{Stefan Wallner}
\affiliation{Max Planck Institute for Physics, 85748 Garching, Germany}

\begin{abstract}
It is long known that interference effects play an important role in understanding the shape of the $\pi^+\pi^-$ spectrum of resonances near the threshold. In this manuscript, we investigate the role of the \rhoomega interference in the study of semileptonic $B \to \pi^+ \pi^- \ell \bar \nu_\ell$ decays. We determine for the first time the strong phase difference between $B \to \rho^0 \ell \bar \nu_\ell$ and $B \to \omega \ell \bar \nu_\ell$ from a recent Belle measurement of the $m_{\pi\pi}$ spectrum of $B \to \pi^+ \pi^- \ell \bar \nu_\ell$. We find \PhaseomrhoFit and extract the branching fraction of \BFrhoFit.  In addition, we set a limit on the $S$-wave component within an $m_{\pi\pi}$ window ranging from $2 m_\pi$ to  $1.02  \, \mathrm{GeV}$ of \BFpipiFit. We also determine the absolute value of the Cabibbo-Kobayashi-Maskawa matrix element of \mbox{$|V_{ub}|_{\rhoomega} = \left( 3.03^{+0.49}_{-0.44} \right) \times 10^{-3}$}, which takes into account the $\rhoomega$ interference. 
\end{abstract}

\maketitle

\section{Introduction}

Determinations of exclusive values of the absolute value of the Cabibbo-Kobayashi-Maskawa matrix element $V_{ub}$ are pre-dominantly carried out using $B \to \pi \ell \bar \nu_\ell$~\cite{HFLAV:2022pwe}, $\Lambda_b \to p \mu \bar \nu_\mu$~\cite{Aaij:2015bfa}, or $B_s \to K \, \mu \bar \nu_\mu$~\cite{LHCb:2020ist} decays.
Determinations using decays $B \to \rho \ell \bar \nu_\ell$, $B \to \omega \ell \bar \nu_\ell$, or higher uncharmed resonances received less attention due to the lack of reliable lattice QCD (LQCD) calculations to predict the corresponding form factors. Here $\rho$ and $\omega$ are referring to the $\rho(770)$ and $\omega(782)$, respectively. Ref.~\cite{Bernlochner:2021rel} provides a world average of
\begin{align} \label{eq:Vub_rho}
 |V_{ub}|_{\,\rho} = \left( 2.96 \pm 0.29 \right) \times 10^{-3} \, , \\
   |V_{ub}|_{\,\omega} = \left( 2.99 \pm 0.35 \right) \times 10^{-3} \, ,
\end{align} 
from combining the available measured differential spectra of $B \to \rho \ell \bar \nu_\ell$ and $B \to \omega \ell \bar \nu_\ell$ decays and using light-cone sum rule (LCSR) calculations of Ref.~\cite{Bharucha:2015bzk} for the form factors.
The resulting values for $|V_{ub}|$ are compatible with each other, but systematically lower than, \eg, the determination from $B \to \pi \ell \bar \nu_\ell$ of Ref.~\cite{HFLAV:2022pwe}
\begin{align}
  |V_{ub}|_{\,\pi} = \left( 3.70 \pm 0.16  \right) \times 10^{-3} \, , 
\end{align}
by about 1.8 or 2.2 standard deviations, respectively. Note that the recent re-analysis of Ref.~\cite{Leljak:2023gna} finds a smaller disagreement. Determinations of $B \to \rho \ell \bar \nu_\ell$ focus both on $\rho^+$ and $\rho^0$ decays into two pions, whereas $B \to \omega \ell \bar \nu_\ell$ focuses on $\omega \to \pi^-\pi^+\pi^0$ or $\omega \to \pi^0 \, \gamma$ decays, cf. measurements published by \babar\ and Belle in Refs~\cite{Sibidanov:2013rkk,delAmoSanchez:2010af,Lees:2012mq}.
The available measurements assume a Breit-Wigner shape for the dynamic amplitude of both resonances. 
Also, they rely on Monte Carlo (MC) simulations to subtract cocktails of resonant and non-resonant $B \to X_u \ell \bar \nu_\ell$ decays.
The size of these contributions though are known to differ depending on the assumptions on the underlying $b \to u \ell \bar \nu_\ell$ MC cocktail or methodology.
Using a so-called ``hybrid'' approach, as originally suggested in Ref.~\cite{PhysRevD.41.1496} and implemented in \eg Refs.~\cite{Belle:2021eni, Belle:2019iji, BaBar:2011xxm}, results in different background estimates as alternative approaches, used to mix exclusive and inclusive $b \to u \ell \bar \nu_\ell$ predictions, as used \eg by Ref.~\cite{Sibidanov:2013rkk}. Both approaches rely on combining simulated decays into known narrow resonances (typically $B \to \{\pi, \rho, \omega, \eta, \eta' \} \ell \bar \nu_\ell$) with scaled predictions from inclusive $B \to X_u \ell \bar \nu_\ell$ calculations, which are hadronized using \texttt{Pythia}~\cite{Sjostrand:1993yb}. None of the state-of-the-art approaches do, however, take into account interference effects. 

To avoid the difficulties to reliably subtract other $b \to u \ell \bar \nu_\ell$ processes that decay into two pions, Ref.~\cite{Belle:2020xgu} measured the $B \to \pi^+ \pi^- \ell \bar \nu_\ell$ process without isolating explicit resonances. The measurement is unfolded from detector effects and reports differential branching fractions as a function of the invariant mass of the di-pion system $m_{\pi\pi}$, the four-momentum transfer squared $q^2$, and in the two dimensions of $q^2: m_{\pi\pi}$. 

\begin{figure}[t!]
\begin{center}
\vspace{8ex}
\includegraphics[width=.5\textwidth]{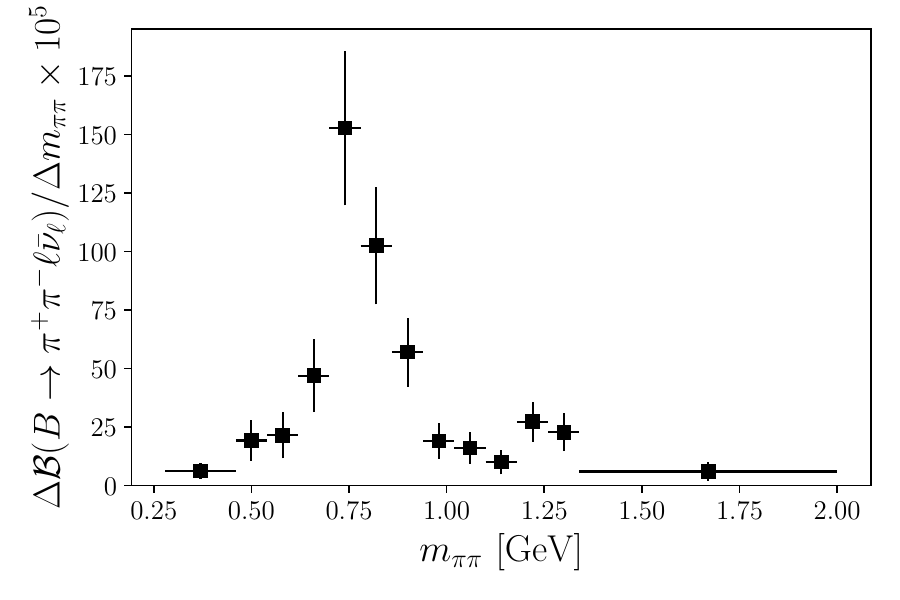}
\caption{ Measured $B \to \pi^+ \pi^- \ell \bar \nu_\ell$ spectrum from Ref.~\cite{Belle:2020xgu}. }
\label{fig:cesars_measurement}
\vspace{-4ex}
\end{center}
\end{figure}

Figure~\ref{fig:cesars_measurement} shows the measured $m_{\pi\pi}$ spectrum ranging from the $2 \pi$ threshold up to \SI{2}{\GeV}. The $\rho$  peak is clearly visible, with a hint of a contribution from the $f_2(1270) \to \pi \pi$ decay around \SI{1.2}{\GeV}. The $m_{\pi\pi}$ region below \SI{0.5}{\GeV} shows enhancements, which might be caused by $\pi \pi$ $S$-wave contributions. 
The shape of the mass spectrum near \mbox{$m_{\pi\pi }\simeq \SI{0.77}{\GeV}$} is strongly affected by the interference of the dominant $\rho$ amplitude with the small contribution of $\omega$ amplitude decaying into $\pi^-\pi^+$.

This seems counter-intuitive at first: the $B \to \rho^0 \ell \bar \nu_\ell$ branching fraction is two orders of magnitude larger than the $B \to \omega ( \to 2 \pi) \, \ell \bar \nu_\ell$ branching fraction~\cite{Bernlochner:2021rel}:
\begin{align} \label{eq:BFrho}
 \mathcal{B} \left( B \to \rho^0 \, \ell \bar \nu_\ell \right) & = \left( 1.35 \pm 0.12 \right) \times 10^{-4} \, , \\ \label{eq:BFom}
 \mathcal{B} \left( B \to \omega ( \to 2 \pi) \, \ell \bar \nu_\ell \right) &= \left( 0.017 \pm 0.002 \right) \times 10^{-4} \, ,
\end{align}
with $\mathcal{B}\left( \omega \to 2\pi  \right) = \left( 1.53 \pm 0.12 \right) \times 10^{-2}$~\cite{Workman:2022ynf}. 

However, as we will see, the interference between both amplitudes distorts the $m_{\pi\pi}$ spectrum with respect to a pure $\rho$ decay. This effect is also observed in a multitude of other processes, such as in $e^+ e^- \to \pi^+ \pi^- (\gamma)$~\cite{BaBar:2012bdw, SND:2020nwa}, in the photoproduction of $\rho$ mesons with gold-gold~\cite{STAR:2017enh} or proton-lead collisions~\cite{CMS:2019awk}, or in the invariant mass spectrum of $e^+e^-$ pairs photoproduced from nuclear targets~\cite{Quinn:1970za}.

The remainder of this manuscript will discuss how all existing measurements of $B \to \rho^0 \ell \bar \nu_\ell$ are affected by interference effects of the $\rho$ signal with $\omega$ contributions. We first recapitulate how different parameterization choices for the dynamic amplitude of the $\rho$ affect its $m_{\pi\pi}$ line shape and peak position. Then we will discuss the formalism to incorporate the \rhoomega interference, and determine both branching fractions and the difference of the strong phases of the amplitudes by analyzing the $m_{\pi\pi}$ spectrum of Ref.~\cite{Belle:2020xgu}. Finally, we set a limit to possible additional S-wave $B \to \pi^+ \pi^- \ell \bar \nu_\ell$ contributions in a mass-window around the $\rho$ resonance.

\begin{figure*}[t!]
\begin{center}
\includegraphics[width=.49\textwidth]{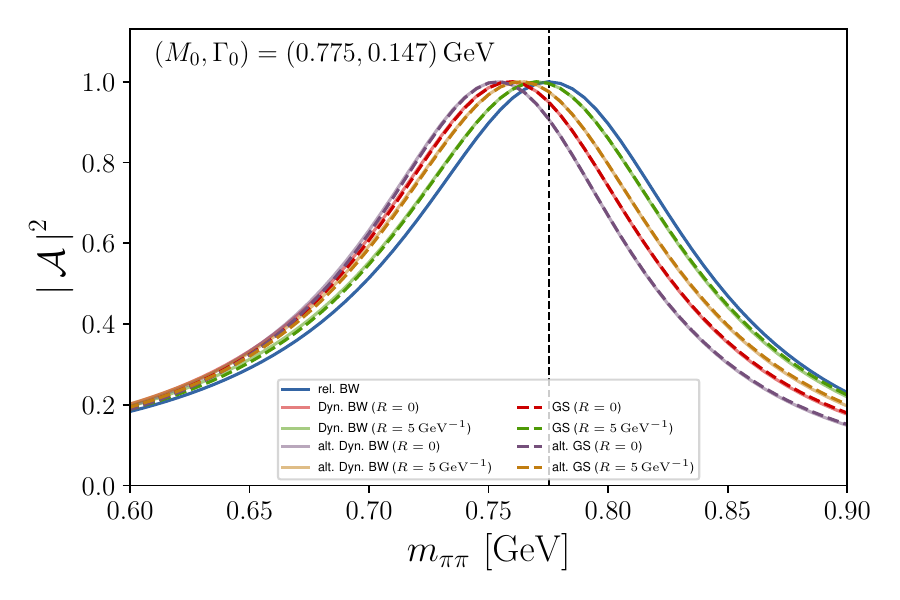}
\includegraphics[width=.49\textwidth]{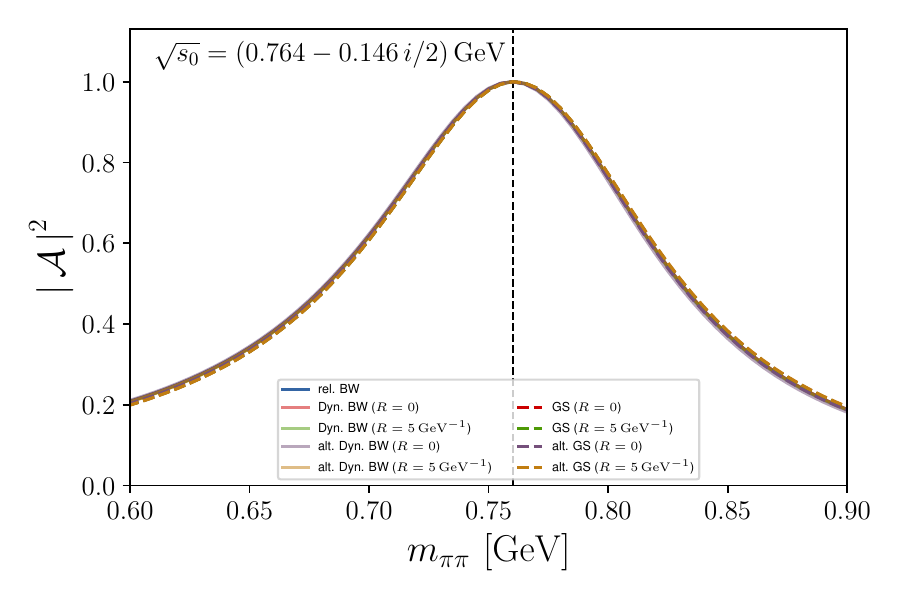}
\caption{ (Left) Line shapes $|\mathcal A(s)|^2$ for different parameterizations using the same of $M_0$ and $\Gamma_0$ parameters. For details of the parameterizations see text. All curves were normalized to their respective modes. (Right) The same comparison, when the parameters of the different parameterizations are chosen such that the parameterizations yield the same pole of $\sqrt{s_0} = \left(0.764 - 0.146 i / 2 \right) \, \mathrm{GeV}$ in the complex $s$ plane.}
\label{fig:line_shape_salad}
\vspace{-4ex}
\end{center}
\end{figure*}

\begin{table*}[ht!]%
\renewcommand{\arraystretch}{1.3}
\centering%
	\caption{A non-exhaustive list of measurements of the $\rho(770)$ parameters using different parameterizations, as extracted from Ref.~\cite{Workman:2022ynf}. If the quoted value for $R$ has an uncertainty, it was fitted to the data, otherwise it was fixed to the quoted value. An $R$ value of \SI{0}{\femto\meter\per\hbar\clight} means that no Blatt Weisskopf barrier factors were used to parameterize the dynamic amplitude. 	
 }%
	\label{tab:rhoMeasurements}%
	\vspace{1ex}
	\begin{tabular}{l|ll|llll|ll}
\toprule
\hline\hline
		                                             & Process                         & Experiment   & $M_0$   [\si{\MeV}]                      & $\Gamma_0$  [\si{\MeV}]                  & $R$  [\si{\femto\meter\per \hbar\clight}] & $R$   [\si{\per\GeV}]                                 & Eq.    & Ref.                                                         \\
                                               \hline
		\midrule
		\multirow{4}{2cm}{Neutral only $e^+ e^-$}      & $e^+ e^- \to \pi^+ \pi^-$        & SND          & $775.3 \pm 0.5 \pm 0.6$      & $145.6\pm0.6\pm 0.8$         & 0 &0                                   & \eqref{eq:dynBW}    & \cite{SND:2020nwa}                                                \\
		                                             & $e^+ e^- \to \pi^+ \pi^-$        & CMD2         & $775.65\pm 0.64\pm 0.50$     & $143.85\pm 1.33\pm 0.80$     & 0 &0                                   & \eqref{eq:GS}       & \cite{Akhmetshin2007}                                             \\
		                                             & $e^+ e^- \to \pi^+ \pi^-\gamma$  & BaBar        & $775.02\pm0.35$              & $149.59\pm0.67$              & 0 &0                                   & \eqref{eq:GS}       & \cite{Lees2012}                                                   \\
		                                             & $\phi \to \pi^+ \pi^- \pi^0$     & KLOE         & $775.9\pm0.5\pm0.5$          & $147.3\pm1.5\pm0.7$          & 0   &0                                 & \eqref{eq:dynBW}    & \cite{KLOE:2003kas}                                               \\
		\hline
		\multirow{3}{2cm}{Charged only $\tau$ decays } & $\tau^- \to \pi^-\pi^0 \nu_\tau$ & Belle        & $774.6\pm0.2\pm0.5$          & $148.1\pm0.4\pm1.7$          & 0 &0                                   & \eqref{eq:GS}       & \cite{Belle:2008xpe}                                              \\
		                                             & $\tau^- \to \pi^-\pi^0 \nu_\tau$ & ALEPH        & $775.5\pm0.7$                & $149.0\pm1.2$                & 0    &0                                & \eqref{eq:GSalt}    & \cite{ALEPH:2005qgp}                                              \\
		                                             & $\tau^- \to \pi^-\pi^0 \nu_\tau$ & CLEO2        & $775.1\pm1.1\pm0.5$          & $150.4\pm1.4\pm1.4$          & 0  &0                                  & \eqref{eq:GS}       & \cite{CLEO:1999dln}                                               \\
		\hline
		\multirow{3}{2cm}{Charged only hadroproduced}  & $\pi^- Cu \to \pi^-\pi^0 Cu$     & SPEC         & $767\pm3$                    & $155\pm11$                   & 0.48    & 2.4                             & \eqref{eq:dynBWalt} & \cite{Capraro:1987rp}                                             \\
		                                             & $\pi^+ A \to  \pi^+\pi^0 A$      & SPEC         & $771\pm4$                    & $150\pm5$                    & 0.47    & 2.4                             & \eqref{eq:dynBWalt} & \cite{Huston:1986wi}                                              \\
		                                             & $\pi p \to \pi \pi N$            & various      & $766.8\pm1.5$                & $148.2\pm4.1$                & $0.33\pm0.02$ & $1.67\pm0.10$                       & \eqref{eq:dynBW}    & \cite{Pisut:1968zza}                                              \\
		\hline
		Mixed other                                  & $p\bar p \to \pi^+\pi^-\pi^0$    & Cryst. Barr. & $763.0\pm0.3\pm1.2$          & $149.5\pm1.3$                & 1.0  & 5.0                                & \eqref{eq:dynBW}    & \cite{CrystalBarrel:1997mmp}                                      \\
		\hline
		\multirow{4}{2.2cm}{Neutral only photoproduced}  & $e p \to e \pi^+\pi^- p$         & H1           & $770.8\pm1.3 ^{+2.3}_{-2.4}$ & $151.3\pm 2.2^{+1.6}_{-2.8}$ & 0    & 0                                & \eqref{eq:dynBW}    & \cite{H1:2020lzc}                                                 \\
		                                             & $e p \to e \pi^+\pi^- p$         & Zeus         & $771\pm2^{+2}_{-1}$          & $155\pm5\pm2$                & 0  & 0                                  & \eqref{eq:dynBW}    & \cite{ZEUS:2011tzw}                                               \\
		% I am not 100% sure, actually they give another formular in the paper, but cite this one from Kuen..?
		                                             & $\gamma p \to \pi^- \pi^+ X$     & Zeus         & $770\pm2\pm1$                & $146\pm3\pm13$               & 0  & 0                                  & \eqref{eq:dynBWalt} & \cite{ZEUS:1997rof}                                               \\
		                                             & $\gamma p \to e^+ e^- p$         & CNTR         & $767.6+\pm2.7$               & $150.9\pm3.0$                & $0.01\pm0.10$   & $0.05\pm0.52$                     & \eqref{eq:dynBW}    & \cite{Bartalucci:1977cp}                                          \\
		\hline
		\multirow{1}{2.0cm}{Neutral other}           & $p \pi^+ \to \pi^+ \pi^- + X$    & HBC          & $768\pm1$                    & $154\pm2$                    & 0  & 0                                  & \eqref{eq:dynBWalt} & \cite{Aachen-Berlin-Bonn-CERN-Cracow-Heidelberg-Warsaw:1975nmd}   \\
		\bottomrule
		%                                            &                                  &              &                              &                              &                                      &                     &                                                                 &
  \hline\hline
	\end{tabular}%
\end{table*}

\section{The many shapes of the $\rho$}\label{sec:shapes_of_rho}

There exist a large number of parameterizations to describe the dynamic amplitude $\mathcal{A}$ of the $\rho(770)$ resonance, and one needs to be careful when choosing a nominal mass $M_0$ and width $\Gamma_0$ from previously reported values. A non-exhaustive list of parameterizations with measured nominal masses and widths is given in Table~\ref{tab:rhoMeasurements}. All parameterizations can be cast into a common form of
\begin{align} \label{eq:gen_form}
    \mathcal{A}(s) = \frac{1}{ M_0^2 - s + f(s) + i M_0 \, \Gamma(s)} \, ,
\end{align}
with  $s = m_{\pi\pi}^2$ and their difference is expressed by the parametrizations for $f(s)$ and $\Gamma(s)$. The simplest choice assumes 
\begin{align}\label{eq:BW}
 f(s) = 0 \, , \qquad \Gamma(s) = \Gamma_0 \, ,
\end{align}
resulting in a {\it fixed width relativistic Breit-Wigner amplitude}. The assumption of $\Gamma(s)$ being constant is only a valid approximation, if the resonance position is far away from the opening of the nearest decay channels. The latter is often expressed as a condition of
\begin{align}
 2 \left(  M_0 - \sqrt{s'} \right) / \Gamma \gg 1 \, .
\end{align}
For the $\pi^+ \pi^-$ channel with $\sqrt{s'} = 2 m_\pi$, we find
\begin{align}
  2 \left(  M_0 - 2 m_\pi \right) / \Gamma \approx 6.7. 
\end{align}
This value might raise some concerns, that the above condition is at best not fully fulfilled for the $\rho$ and possible deviations should be explored. 

The $s$-dependence on the width is often taken into account using the so-called {\it dynamic-width Breit-Wigner amplitude}, which for a single decay channel reads
\begin{align} \label{eq:dynBW} 
  f(s) & = 0 \, , \quad   \Gamma(s) = \Gamma_{\mathrm{BW}}(s) =  \Gamma_0 \frac{M_0}{\sqrt{s}} \frac{q^3}{q_0^3} \, \frac{\mathcal{F}_1(R \, q)}{ \mathcal{F}_1(R \, q_0) } \, .
\end{align}
Here, $q = q(s, m_\pi, m_\pi)$ denotes the two-body break-up momentum 
\begin{align}
 q(s,M_1,M_2) = \frac{ \lambda^{1/2} (s, M_1^2, M_2^2) }{2 \, \sqrt{s}} \, ,
\end{align}
with $\lambda$ being the K\"all\'en function~\cite{Kallen:1964lxa}, and \mbox{$q_0 = q(M_0, m_\pi, m_\pi)$}. Further, 
\begin{align}\label{eq:BW_factor}
 \mathcal{F}_1( R q ) & = 1 / \left( 1 + \left( R q \right)^2 \right) \, ,
\end{align}
is the Blatt-Weisskopf~\cite{Blatt:1952ije} factor, with $R$ is a scale factor related to the radius of the strong potential, which determines the barrier of the angular momentum of $R \cdot q$. Note that we dropped an overall normalization factor, that cancels in the ratio of Eq.~\ref{eq:dynBW}. 
An extension of Eq.~\ref{eq:dynBW} is the so-called {\it Gounaris-Sakurai amplitude}~\cite{Gounaris1968} with 
\begin{align} \nonumber
 f(s) & = f_{\mathrm{GS}}(s) = \frac{ \Gamma_0 \, M_0^2 }{ q_0^3}  \left( q^2 \left( h - h_0 \right)  +  \left( M_0^2 - s \right) q_0^2 h'_0 \right)  \, , \\\label{eq:GS}
 \Gamma(s) & = \Gamma_{\mathrm{GS}}(s) = \Gamma_{\mathrm{BW}}(s) \, . 
\end{align}
Here, $h = h(s)$ and $h_0 = h(M_0^2)$ with
\begin{align}
 h(s) = \frac{2}{\pi} \frac{q(s)}{ \sqrt{s} } \ln \left( \frac{\sqrt{s} + 2 q(s)}{ 2 m_\pi}  \right) \, .
\end{align}
and $h'_0 = h_0 \left(  (8 q_0^2)^{-1} - (2 M_0^2)^{-1}  \right) + (2 \pi \, M_0^2)^{-1}$. Eq.~\ref{eq:GS} is often used, especially when determining the pole position of the amplitude. 

The energy dependence of the dynamic width in Eq.~\ref{eq:dynBW} is not unique and other choices exist~\cite{Pisut:1968zza}: one can introduce an additional factor of $\sqrt{s}/M_0$ that modifies Eqs.~\ref{eq:dynBW} and \ref{eq:GS} such that
\begin{align} \label{eq:dynBWalt}
 f(s) = 0 \, ,\quad \Gamma(s) = \Gamma(s)_{\mathrm{aBW}} = \frac{\sqrt{s}}{M_0} \, \Gamma(s)_{\mathrm{BW}}\, ,
\end{align}
and
\begin{align} \label{eq:GSalt}
 f(s) & = f_{\mathrm{aGS}} = f_{\mathrm{GS}}(s)\, ,\quad \Gamma(s) = \Gamma(s)_{\mathrm{aGS}} = \frac{\sqrt{s}}{M_0} \, \Gamma(s)_{\mathrm{GS}}\, ,
\end{align}

To widen the amount of possible parameterizations even further, other authors omit the Blatt-Weisskopf barrier factors Eq.~\ref{eq:BW_factor}, which is equivalent to choosing $R = \SI{0}{\per\GeV}$.

With these different choices at hand, we will now investigate the various resulting line shapes $|\mathcal A|^2$ using the same choice for $M_0$ and $\Gamma_0$ . We use $M_0 = \SI{0.775}{\GeV}$ and $\Gamma_0 = \SI{0.147}{\GeV}$ and either set $R = 0$ or $R = \SI{5}{\per\GeV}$. The line shapes are shown in Figure~\ref{fig:line_shape_salad}. We note that the dynamic-width Breit-Wigner Eq.~\ref{eq:dynBW} and the Gounaris-Sakurai amplitude Eq.~\ref{eq:GS} give very similar shapes, with some minor differences in the tails of the resonance peak. The value of the scale parameter $R$ impacts the position of the peak of the line shape, resulting in a positive shift of about \SI{10}{\MeV} when going from $R = 0 \to \SI{5}{\per\GeV}$. Using the alternative $s$-dependencies (Eqs.~\ref{eq:dynBWalt} and \ref{eq:GSalt}) results in a negative shift of about \SI{5}{\MeV} of the peak position compared to the nominal parameterizations.

The observed shifts in the peak position for the different parameterizations are, however, not a physical property of the $\rho$ resonance. They are an artefact of the parameterizations. The universal physical properties of a resonance are described by the position of the pole of the corresponding amplitude in the complex $s$-plane, and $M_0$ and $\Gamma_0$ depend on the choice of parameterization. Requiring the same pole position of $\sqrt{s_0} = \left(0.764 - 0.146 i / 2 \right) \, \mathrm{GeV}$~\cite{Garcia-Martin:2011nna,Workman:2022ynf} for each parameterization by solving for the corresponding values of $M_0$ and $\Gamma_0$ results in nearly identical line shapes as shown in figure~\ref{fig:line_shape_salad} (right).

In order to model the measured $m_{\pi\pi}$ spectrum in $B \to \pi \pi \ell \bar \nu_\ell$ decays, we multiply the line shape with two additional factors: the $\pi \pi \ell \bar \nu_\ell$ phase space $\Phi(s)$ and the angular-momentum barrier factor $\mathcal{B}_F(s, R)$~\cite{VonHippel:1972fg}. The $\mathcal{B}_F$ factor models centrifugal-barrier effects that distort the line shape. 
We approximate the $m_{\pi\pi}$ dependence of the phase-space by the product of the two-body break-up momenta of the $B$ decay to the $\pi\pi$ and $\ell \bar \nu_\ell$ systems and the decay of the $\pi\pi$ system,
\begin{align} \label{eq:PS}
    \Phi(s) \approx q(M_B^2, \sqrt{s}, m_{\ell \overline \nu_\ell}) \cdot q(s, m_\pi, m_\pi) \,  .
\end{align}
We choose a constant invariant mass for the $\ell \bar \nu_\ell$-system of $m_{\ell \overline \nu_\ell} = \SI{3.6}{\GeV}$, determined from a fit to a Monte Carlo sample that is uniformly distributed in the phase-space of the studies process. Eq.~\ref{eq:PS} provides an accurate description for the region of $m_{\pi\pi} < \SI{1}{\GeV}$, which is the relevant range for our analysis. Finally, the barrier factor is given by
\begin{align}\label{eq:BF}
\mathcal{B}_{F}(s, R) = q^2 \cdot \mathcal{F}_1\left(R q\right)  \, ,
\end{align}
with $q =  q(s, m_\pi, m_\pi) $. 

\begin{figure}[t!]
\begin{center}
\includegraphics[width=.5\textwidth]{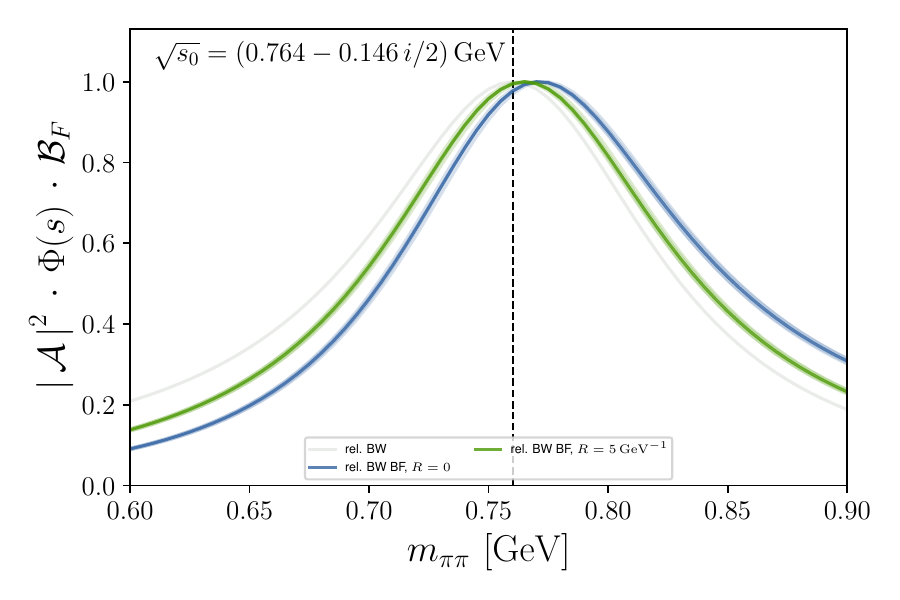}
\caption{Model for the $m_{\pi\pi}$ spectrum in $B \to \pi \pi \ell \bar \nu_\ell$ decays using the relativistic Breit-Wigner amplitude The blue curve shows the model for $R = 0$, whereas the green curve for $R = \SI{5}{\per\GeV}$. The gray curve shows the line shape, i.e. the model without the phase-space and barrier factors applied for comparison. For details see text.}
\label{fig:final_rho_line_shape}
\vspace{-4ex}
\end{center}
\end{figure}

Figure~\ref{fig:final_rho_line_shape} shows our model for the $m_{\pi\pi}$ spectrum using the relativistic Breit-Wigner Eq.~\ref{eq:BW} for $\sqrt{s_0} = \left( 0.764 - 0.146 i /2 \right) \, \mathrm{GeV}$ (grey curve) without any additional factors applied. The blue and green curve show the line shapes with both factors applied for $R = 0$ and $R = \SI{5}{\per\GeV}$. Note that the pole position in the complex plane is not affected by $\Phi$ and $\mathcal{B}_F$, but the peak position in the $m_{\pi\pi}$ spectrum is shifted and hence sensitive to the choice of the momentum scale parameter $R$. The blue and green shaded bands represent the uncertainties on $M_0$ and $\Gamma_0$, propagated from the uncertainties on the real and imaginary part of $\sqrt{s_0}$ from the global fit in Ref.~\cite{Garcia-Martin:2011nna}, which are small and barely visible in Figure \ref{fig:final_rho_line_shape}.

\begin{figure}[t!]
\begin{center}
\includegraphics[width=.5\textwidth]{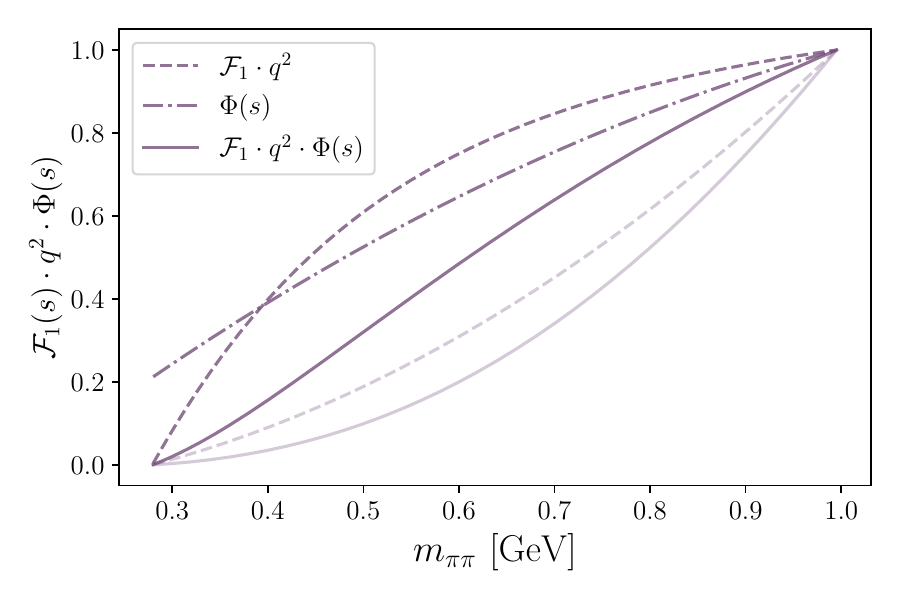}
\caption{ The impact of the phase space and barrier factor, that lead to the asymmetry of the line shape as a function of $m_{\pi\pi}$ are shown for $R = \SI{5}{\per\GeV}$ (plum) and $R = 0$ (light plum). All curves are normalized to their mode.}
\label{fig:phi_BF_factors}
\vspace{-4ex}
\end{center}
\end{figure}

Figure~\ref{fig:phi_BF_factors} shows the functional dependence of $\Phi$ and $\mathcal{B}_F$. For $R = 0$ the Blatt-Weisskopf factor is constant and the phase space and $q^2$ factors both increase as a function of $m_{\pi\pi}$. This results in an asymmetry around the $\rho$ peak of the line shape. If $R \neq 0$, the Blatt-Weisskopf factor decreases as a function of $m_{\pi\pi}$, reducing the size of this asymmetry. 

\section{When $\rho^0$ meets $\omega$ interference ensues} \label{sec:interference}

We now turn our attention to the \rhoomega interference, which becomes visible due to the isospin breaking decay of the $\omega$ meson into two pions~\cite{Gourdin:1969kc}. The origin of this effect is that the physical observable $\rho_{\mathrm{ph}}$ and $\omega_{\mathrm{ph}}$ states are a superposition of the pure $\rho$ and $\omega$ isospin states,
\begin{align}
| \rho_{\mathrm{ph}} \rangle =  | \rho \rangle - \epsilon  | \omega \rangle  \, , \\
| \omega_{\mathrm{ph}} \rangle =  \epsilon | \rho \rangle + | \omega \rangle  \, , 
\end{align}
with $\epsilon$ the electromagnetic admixture. 

This interference can be formally introduced using a complex-valued mixing matrix~\cite{Goldhaber:1969dp,Coleman:1964ac,Harte:1964zz,PERensing} 
\begin{align}
 \mathcal{M} = \left( \begin{matrix} m_\rho^2 - i m_\rho \Gamma_\rho & - \delta\left( m_\rho + m_\omega \right)  \\ - \delta \left( m_\rho + m_\omega \right) & m_\omega^2 - i m_\omega \Gamma_\omega  \end{matrix}  \right) \, ,
\end{align}
with the strength of the electromagnetic mixing expressed as a complex-valued parameter $\delta \sim \epsilon$. The total \rhoomega amplitude is then given by
\begin{align} \label{eq:m_intf}
 \mathcal{A}_{\rhoomega}  =  \left( \mathcal{P}_{\rho}, \mathcal{P}_{\omega}  \right)  \left\{ \mathcal{M} - s \bold{1} \right\}^{-1} \left( \begin{matrix} \mathcal{D}_{\rho} \\ \mathcal{D}_{\omega}  \end{matrix} \right) \, ,
\end{align}
with $\mathcal{P}$ and $\mathcal{D}$ denoting the production and decay amplitudes of the pure isospin states $\rho$ and $\omega$, and $\bold{1} $ the $2\times 2$ unit matrix. 

Neglecting the small direct decay amplitude of $\omega \to \pi \pi$, Eq.~\ref{eq:m_intf} can be simplified to~\cite{Back:2017zqt}
\begin{align} \label{eq:A_r_w}
 \mathcal{A}_{\rhoomega} = \mathcal{A}_{\rho} \left( \frac{1 + \mathcal{A}_{\omega} \, \Delta \, |B| e^{i \phi_{\rhoomega}}}{1 - \Delta^2 \mathcal{A}_{\rho} \mathcal{A}_{\omega}}  \right) \, ,
\end{align}
with $\mathcal A_{\rho/\omega}$ denoting the amplitude of $\rho$ or $\omega$, respectively, and $\Delta = \delta \left( m_\rho + m_\omega \right)$. Further, 
\begin{align}
 B = \mathcal{P}_{\omega} / \mathcal{P}_{\rho_0} = |B| e^{i \phi_{\rhoomega}} \, ,
\end{align}
with  $\phi_{\rhoomega}$ denoting relative strong phase difference between the $\rho$ and $\omega$ production amplitudes and $|B|$ is proportional to the square-root of the production branching fractions of $B \to \rho^0 \, \ell \bar \nu_\ell$ and $B \to \omega \, \ell \bar \nu_\ell$, cf. Appendix~\ref{app:D}.

In contrast to the $\rho$, the $\omega$ is a narrow resonance with a width of $\simeq \SI{8.7}{\MeV}$, and none of the effects discussed in Section~\ref{sec:shapes_of_rho} have any sizeable impact on its line shape. We thus describe $\mathcal{A}_\omega$ using a fixed width relativistic Breit-Wigner amplitude according to Eq.~\ref{eq:BW} with mass and width from Ref.~\cite{Workman:2022ynf}. We explicitly investigated different choices for the $\omega$ parameterization and found their impact to be negligible.

We will use for the electromagnetic mixing the parameters $|\delta| = \left( 2.15 \pm 0.35 \right) \mathrm{MeV}$~\cite{PERensing} and $\arg \delta = 0.22 \pm 0.06$~\cite{CMD-2:2001ski}. The phase and absolute value of $\delta$ can be measured with the $e^+ e^- \to \pi^+ \pi^-$ process: due to the production process via a virtual photon no strong phase difference is introducing an additional phase between the $\rho$ and $\omega$ amplitudes. 
%We use the square root of the semileptonic $\omega$ and $\rho$ production branching fractions (cf. Eqs.~\ref{eq:BFrho} and \ref{eq:BFom}) and determine $|B| = 0.92\pm0.07$.

\begin{figure}[t!]
\begin{center}
\includegraphics[width=.5\textwidth]{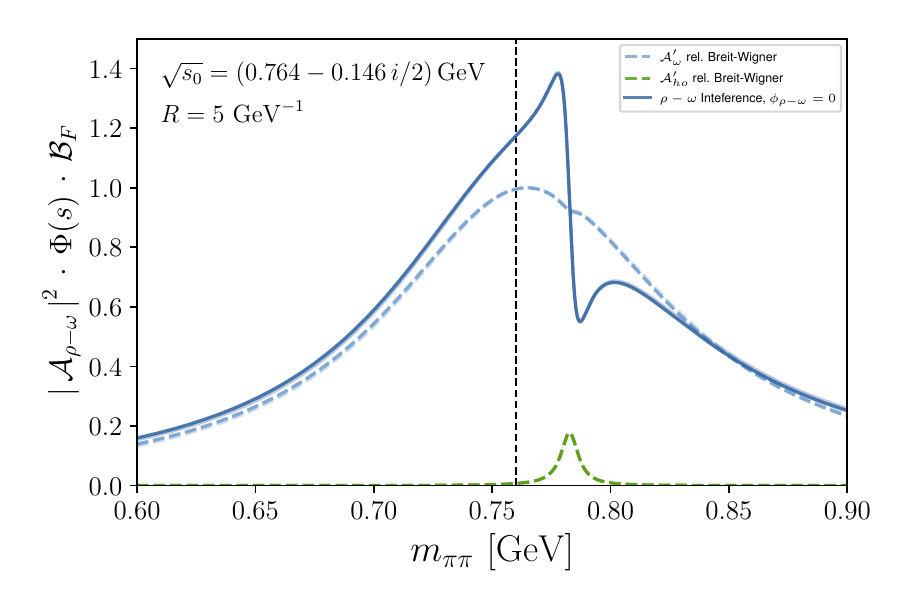}
\caption{ The $m_{\pi\pi}$ spectrum from the interference is shown assuming $\rho$ and $\omega$ are produced with $\phi_{\rhoomega} = 0$ (solid line), using the relativistic Breit-Wigner amplitude with mass and width chosen to possess a pole at $\sqrt{s_0} = (0.764-0.146i/2)\,\si\GeV$ and $R = 5 \, \mathrm{GeV}^{-1}$. The shaded light blue lines show the interference prediction using the other parameterizations which mass and width chosen to yield the same pole position. The dashed lines show the spectrum for the pure $\rho$ (blue) and $\omega$ (green) contributions.}
\label{fig:line_shape}
\vspace{-4ex}
\end{center}
\end{figure}

Figure~\ref{fig:line_shape} illustrates the impact of the interference on the $\pi\pi$ spectrum if both $\rho$ and $\omega$ are produced fully coherently with $\phi_{\rhoomega} = 0$. The $\pi\pi$ spectra of the pure $\rho$ and $\omega$ contributions, defined as $\left|\mathcal{A}_\rho'\right|^2 = \left|\mathcal{A}_\rho / \left( 1 - \Delta^2 A_\rho A_\omega \right)\right|^2$ and $\left|A_{\omega}'\right|^2 = \left|\mathcal{A}_\rho \mathcal{A}_\omega  \Delta |B| / \left( 1 - \Delta^2 A_\rho A_\omega \right)\right|^2$, are shown as dashed curves. The small crest near the $\omega$ mass on $\mathcal{A}_\rho'$ stems from the $\left( 1 - \Delta^2 A_\rho A_\omega \right)$ term. Due to the interference, the resulting $\pi\pi$ line shape is strongly distorted near the $\omega$ mass, resulting in a cusp. 

\begin{figure}[t!]
\begin{center}
\includegraphics[width=.5\textwidth]{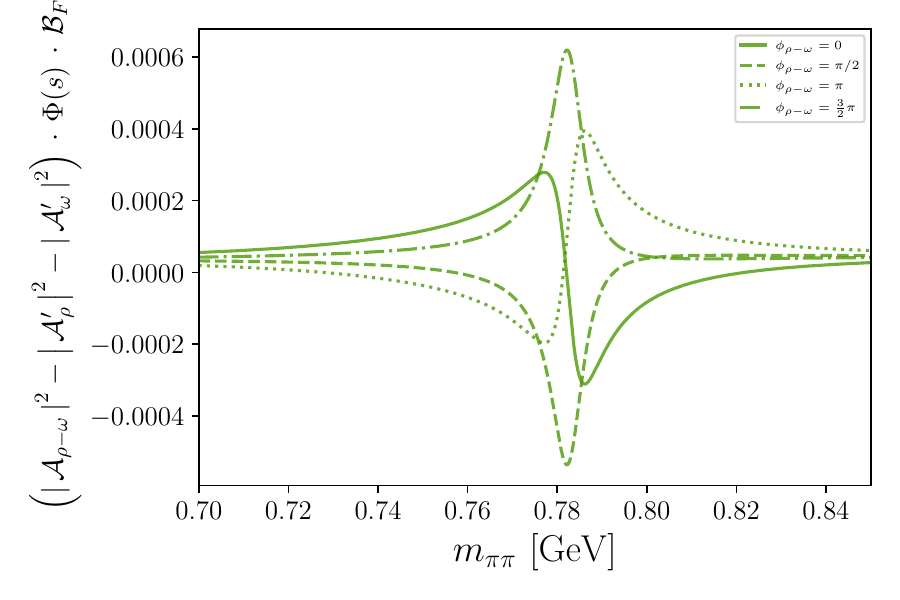}
\caption{ The distortion of the $m_{\pi\pi}$ spectrum from interference is shown for four different values of $\phi_{\rhoomega}$.}
\label{fig:interference}
\vspace{-4ex}
\end{center}
\end{figure}

Figure~\ref{fig:interference} shows the distortion with respect to the incoherent sum of the pure $\rho$ and $\omega$ contributions for four different choices of $\phi_{\rhoomega} \in [0, \frac{\pi}{2}, \pi, \frac{3}{2}\pi]$. If both states are produced with a relative phase difference of $\pi$ instead of $0$, the distortion of the spectrum changes sign, resulting in a depletion below the $\omega$ mass and an enhancement above. Fractional phase shifts in $\pi$ of a half or three halves result in an enhancement or attenuation of the total signal due to constructive or destructive interference. 

\section{$S$-wave and Isobar Model}
\label{sec:swave}

In addition to the $\rho$ and $\omega$ vector mesons, we also include a \mbox{$B \to [\pi \pi]_S \ell \bar \nu_\ell$} $S$-wave contribution to describe the low mass $m_{\pi\pi}$ spectrum. The importance to study such a contribution in the context of $B \to \rho^0 \ell \bar \nu_\ell$ was pointed out in Ref.~\cite{Kang:2013jaa}. The $S$-wave can be calculated in a model independent way using dispersion theory, using the measured $\pi\pi$ phase shifts and a couple channel treatment for the $K \overline K$ system~\cite{Daub:2015xja}. This requires knowledge of the Omn\'es matrix 
and the pion and kaon form factors at $s = 0$. The resulting line shape can only be obtained numerically and we use values from the authors of Ref.~\cite{Daub:2015xja} provided in Ref.~\cite{cesarthesis}. We also study alternative descriptions of this shape: we implement a simplified model of the interplay of the $f_0(500)$ and $f_0(980)$ resonances used by Ref.~\cite{COMPASS:2015gxz} and based of Ref.~\cite{Au:1986vs}. This model also uses information obtained from $\pi\pi$ elastic scattering data but removes the $f_0(980)$ from the description of the $S$-wave amplitude.  We further carry out fits assuming a uniform phase space distribution according to Eq.~\ref{eq:PS}.

\begin{figure}
\begin{center}
\includegraphics[width=.45\textwidth]{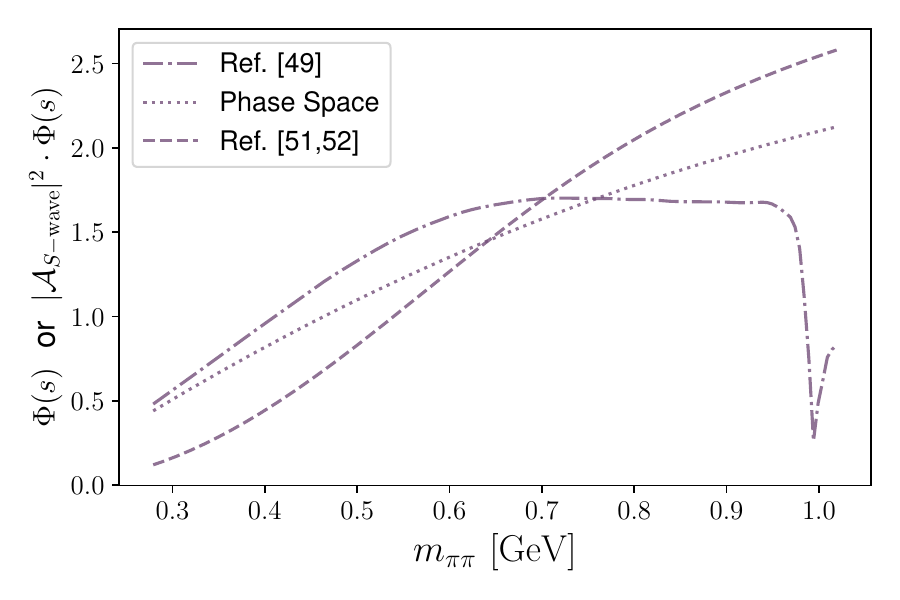}
\vspace{-2ex}
\caption{ The considered parameterizations for the $S$-wave contribution are shown.   }
\label{fig:bkg_modelling}
\vspace{-6ex}
\end{center}
\end{figure}

Figure~\ref{fig:bkg_modelling} compares the predicted $S$-wave mass distribution: the predicted $S$-wave of Ref.~\cite{Daub:2015xja} (dash-dotted curve) enhances the low $m_{\pi\pi}$ region and falls off and produces a cusp around \SI{1}{\GeV}. The prediction of Refs.~\cite{COMPASS:2015gxz,Au:1986vs} (dashed curve) predicts a depletion at low $m_{\pi\pi}$, and then raises steeply. Phase space predicts (dotted curve) a steadily raising distribution, which raises slower than the model of Refs.~\cite{COMPASS:2015gxz,Au:1986vs}. In the following we will use the model of Ref.~\cite{Daub:2015xja} as our default parameterization as it provides the most complete description of the $\pi\pi$ $S$-wave contribution. But the other models result in very similar results and are fully discussed in Appendix~\ref{app:A}.

\begin{table}[b]
\renewcommand{\arraystretch}{1.3}
\caption{ The used external inputs and references are listed. }
\vspace{2ex}
\begin{tabular}{c|c|c}
\hline\hline
   Observable  & Value & Ref.  \\
   \hline
   $ \sqrt{s_0}$ & $\left( 763.7^{+1.7}_{-1.5} - 73.2^{+1.0}_{-1.2} i \right) \, \mathrm{MeV}$ & \cite{Garcia-Martin:2011nna} \\
   $M_\omega$ & $\left( 782.66 \pm 0.13 \right) \, \mathrm{MeV}$ & \cite{Workman:2022ynf} \\
   $\Gamma_\omega$ & $ \left(8.68 \pm 0.13 \right) \, \mathrm{MeV}$ &  \cite{Workman:2022ynf} \\
   $R$ & $5.3^{+0.9}_{-0.7} \, \mathrm{GeV}^{-1}$ & \cite{Workman:2022ynf, CERN-Cracow-Munich:1982jjg}\\
   $|\delta|$ & $\left( 2.15 \pm 0.35 \right) \, \mathrm{MeV}$ & \cite{PERensing} \\
   $\arg{\delta}$ & $0.22 \pm 0.06$ & \cite{CMD-2:2001ski} \\
   $\mathcal{B} \left( B \to \omega ( \to 2 \pi) \, \ell \bar \nu_\ell \right)$  & $\left( 0.017 \pm 0.002 \right) \times 10^{-4} $ & \cite{Bernlochner:2021rel,Workman:2022ynf} \\  
\hline\hline
\end{tabular}
\label{tab:fit_input}
\end{table}

We will study the $m_{\pi\pi}$ spectrum using an isobar model approach~\cite{Fleming:1964zz, Morgan:1968zza, Herndon:1973yn}, describing the full $B \to \pi \pi \ell \bar \nu_\ell $ decay amplitude using the incoherent sum of the \rhoomega contribution and the $S$-wave part.
Treating the $S$-wave incoherently is justified as we only analyze the $m_{\pi\pi}$ spectrum and hence integrate over all of the decay angles of the process. As the angular distributions of the $S$-wave contribution and the $P$-wave \rhoomega contribution are orthogonal, their interference vanishes.
Note that in principle a non-uniform experimental acceptance in the decay angles could break this orthogonality in practice and produce non-vanishing interference distortions in the experimental $m_{\pi\pi}$ spectrum. But such effects need to be studied by the experimental collaborations and are beyond the scope of this paper.

\section{Fit Setup}

We have now assembled all the individual pieces to finally analyze the  $m_{\pi\pi}$ spectrum: We will study its composition using a $\chi^2$ fit of the form 
\begin{align} \label{eq:chi2}
 \chi^2 = \left( \bold{\Gamma}_m - \bold{\Gamma}_p  \right) C^{-1} \left(\bold{\Gamma}_m - \bold{\Gamma}_p  \right) + \sum_k \chi^2_{k} \, ,
\end{align}
with $\left(\bold{\Gamma}_m \right)_i$ and $C$ denoting the measured spectrum in a given $m_{\pi\pi}$ bin $i$  and the statistical and systematic covariance matrix of Ref.~\cite{Belle:2020xgu}.

The prediction of the $m_{\pi\pi}$ line shape is constructed from integrals of the form
\begin{align}
    \begin{aligned}
        \left( \bold{\Gamma}_p \right)_i = \int\limits_{\mathclap{\Delta m_{\pi\pi\, i}}}\mathrm{d} m_{\pi\pi}&\left(  \mathcal{B}_\rho \cdot \left|\mathcal A_\rhoomega\right|^2 \cdot  \Phi \cdot \mathcal B_F \right .\\[-1em]
        &\left. \quad\,\, + \quad \mathcal{B}_{S} \cdot \left|\mathcal A_{S-\mathrm{wave}}\right|^2 \cdot  \Phi  
 \, \, \right) \, .
    \end{aligned} 
\end{align}
The parameters of interest determined by the fit are the \mbox{$B \to \rho^0 \ell \bar \nu_\ell$} branching fraction proportional to $\mathcal{B}_{\rho}$, the strong phase difference $\phi_{\rhoomega}$ (encapsulated in $\mathcal A_\rhoomega$), and the $B \to [\pi \pi]_S \ell \bar \nu_\ell$ $S$-wave contribution proportional to $\mathcal{B}_{S}$. Appendix~\ref{app:D} provides the concrete relations.

Additional parameters, for example the masses and widths of resonances, are constrained to external inputs using symmetric or asymmetric Gaussian constraints with $\sigma_k^\pm$ denoting the upper or lower uncertainty via
\begin{align}
  \chi^2_{k} = \frac{\left( \theta_k^{m} - \theta_k^{p} \right)^2}{\sigma^2_k} \, \text{, with} \, \sigma_k = \bigg\{ \begin{matrix}  \theta_k^{p} > \theta_k^{m}  & \sigma_k = \sigma_k^+ \\
   \theta_k^{p} \leq \theta_k^{m}  & \sigma_k = \sigma_k^- \\
  \end{matrix}  \, .
\end{align}
Here, $\theta_k^{m/p}$ denotes either the external or predicted value for the external input. Table~\ref{tab:fit_input} provides an overview of all external parameters.

We constrain the pole position of the $\rho$ to $\sqrt{s_0} = \left( 0.764 - 0.146 i /2 \right) \, \mathrm{GeV}$ from Ref.~\cite{Garcia-Martin:2011nna}. This is realized by numerically evaluating the pole of the employed parameterization as a function of $M_\rho$ and $\Gamma_\rho$ (and $R$ when appropriate) in each iteration of the fit. The mass and width of the $\omega$ contribution are constrained to $(M_\omega, \Gamma_\omega) = \left(0.783, 0.009 \right) \mathrm{GeV}$~\cite{Workman:2022ynf}. We constrain the momentum scale parameter to $R = 5.3^{+0.9}_{-0.7} \, \mathrm{GeV}^{-1}$ from Ref.~\cite{Workman:2022ynf} based on the determination of Ref.~\cite{CERN-Cracow-Munich:1982jjg} unless stated otherwise. The absolute value and argument of the electromagnetic mixing operator are constrained to the values of Ref.~\cite{PERensing} and \cite{CMD-2:2001ski}. We further constrain the  $B \to \omega ( \to 2 \pi) \, \ell \bar \nu_\ell $ mixing contribution using the branching fractions of Ref.~\cite{Bernlochner:2021rel,Workman:2022ynf}. 

We numerically minimize the $\chi^2$ of Eq.~\ref{eq:chi2} using the \texttt{iMinuit} package~\cite{iminuit}. We profile $\Delta \chi^2 = \chi^2 - \chi^2_{\mathrm{min}}$ with $\chi^2_{\mathrm{min}}$  the minimal value of the $\chi^2$ function to determine the uncertainties of all fit parameters. We further determine numerically the approximate covariance matrix from the second-order partial derivatives of the $\chi^2$ function at the best fit point.

\section{Results}

\begin{figure}
\begin{center}
\includegraphics[width=.5\textwidth]{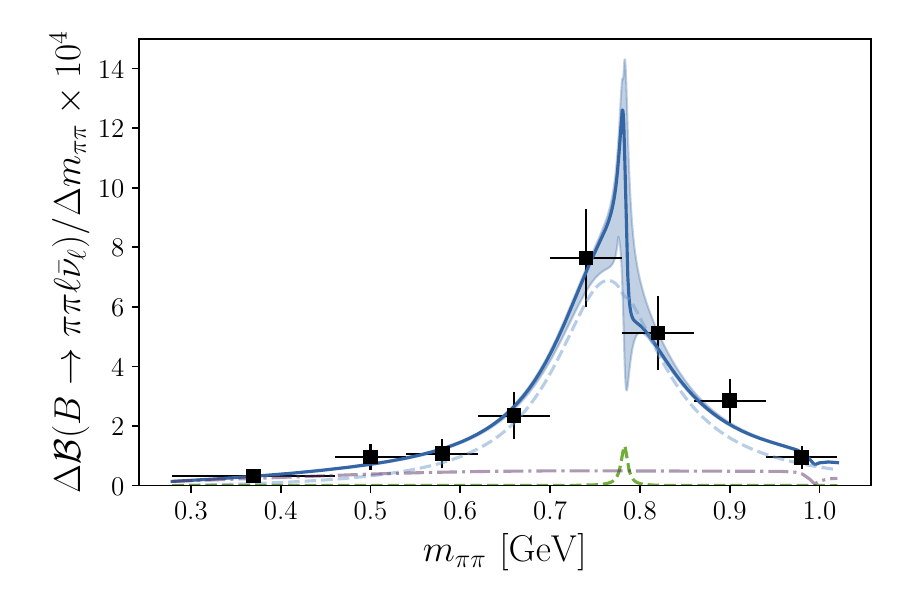}
\vspace{-4ex}
\caption{ Fit result using the relativistic Breit-Wigner to model the $\rho$ amplitude and using the shape of Ref.~\cite{Daub:2015xja} to model the $S$-wave background. The blue line and shaded band show the total model curve and its uncertainty. The $S$-wave contribution is shown as a dash-dotted purple line. The dashed lines show the prediction of the pure $\rho$ (blue) and $\omega$ (green) contributions. }
\label{fig:fit_BW}
\vspace{-4ex}
\end{center}
\end{figure}

\begin{figure}
\begin{center}
\includegraphics[width=.5\textwidth]{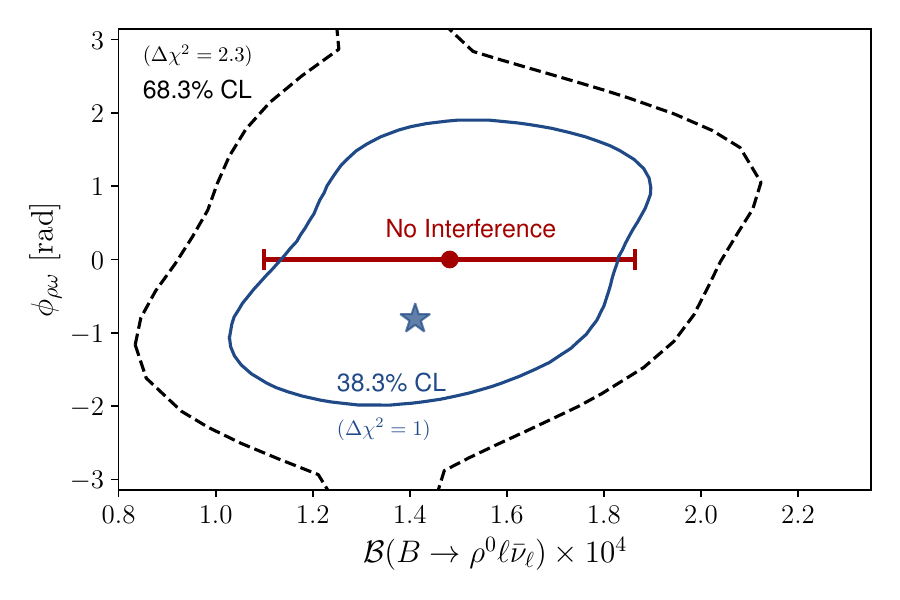}
\caption{ The $\Delta \chi^2 = 1$ (38.3\% CL, blue) and $\Delta \chi^2 =2.3$ (68.3 \% CL, black dashed) contours of \BFrho and $\phi_{\rhoomega}$ are shown for the $S$-wave described using Ref.~\cite{Kang:2013jaa}. The best fit point ($\Delta \chi^2 = 0$) is indicated with a blue star. The red data point shows the result of a fit neglecting the interference effects between $\rho$ and $\omega$.   }
\label{fig:fit_BW_2}
\vspace{-4ex}
\end{center}
\end{figure}

As we demonstrated in Sections~\ref{sec:shapes_of_rho} and \ref{sec:interference} the choice of the $\rho$ parameterization is not important as long as the same physical pole in $\sqrt{s_0}$ is imposed. We thus describe the $\rho$ amplitude with a relativistic Breit-Wigner, imposing all constraints listed in Table~\ref{tab:fit_input}, and describe the $S$-wave contribution with the prediction of Ref.~\cite{Daub:2015xja}. We determine:
\begin{align} \label{eq:res_rho_BW}
 \BFrho & = \BFrhoFitValue  \, ,  \\ \label{eq:res_rho_BW_phase}
 \phi_{\rhoomega} & = \PhaseomrhoFitValue \, .
 \end{align}
Figure~\ref{fig:fit_BW} depicts the result of the fit. No statistically significant contribution of the $S$-wave was found and we determine an upper limit of
 \begin{align} 
 \Delta\mathcal{B}(B \to [\pi^+ \pi^-]_S \, \ell \bar \nu_\ell) & <  \BFpipiFitValue \, ,
 \end{align}
 defined as a partial branching fraction inside the window of $m_{\pi\pi} \in [2 m_\pi, 1.02\, \mathrm{GeV}]$. The correlation matrix between between the three parameters is
 \begin{align}
   C = \left( \begin{matrix} \phantom{+}1.00 & \phantom{+}0.27 & -0.43 \\
                              \phantom{+}0.27 & \phantom{+}1.00 & -0.10 \\
                              -0.43 & -0.10 & \phantom{+}1.00 \end{matrix} \right)  \, .
 \end{align}
 The $S$-wave contribution is $-43\%$ anti-correlated with the $\rho$ branching fraction. The strong phase difference is 27\% correlated with the $\rho$ branching fraction and -10\% anti-correlated with the $S$-wave. The $\chi^2$ of the fit is 2.07 with 5 degrees of freedom, resulting in a $p$-value of 83.9\%. The determined values and uncertainties of all fit parameters are summarized in Table~\ref{tab:fit_results_rho_BW_SW} and Figure~\ref{fig:fit_BW_2} shows the two-dimensional contours for fixed $\Delta \chi^2$ spanned by the $B \to \rho^0 \ell \bar \nu_\ell$ branching fraction and $ \phi_{\rhoomega}$. 
 
 We can also assess the branching fraction of the physical $\rho_\mathrm{ph}$ and $\omega_{\mathrm{ph}}$ states, that decay into two pions, defined as the admixture of the $\rho$ and $\omega$ isospin states. This branching fraction represents the dominant $P$-wave contribution of the low $m_{\pi\pi}$ spectrum and we find
 \begin{align}
 \mathcal{B} (B \to  [\pi^+ \pi^-]_{\rhoomega} \, \ell \bar \nu_\ell ) & = \left( 1.49 \pm 0.38 \right) \times 10^{-4}  \, .
 \end{align}

 Using alternative parameterizations to describe the $\rho$ result in very similar branching fractions and strong phases. A full summary is listed in Table~\ref{tab:fit_results_rho_all_SW}. The variations related to the various parametrizations are no larger than $\simeq 0.01 \times 10^{-4}$ for the $\rho$ branching fraction or $2 \degree$  for the strong phase difference. This is consistent with the shift in the phase dependence from the alternative parameterizations. Particularly the phase of the Gounaris-Sakurai parameterization has a different $s$ dependence, resulting in a smaller value of the strong phase of $\left( -44_{\,-67}^{+150} \right)\degree$. This difference, however, is not relevant and the recovered $\rho$ branching fraction is nearly parameterization independent.
 
By removing the constraint on the momentum scale parameter $R$, a marginally smaller branching fraction of \mbox{$\BFrho  = \left(1.39^{+0.46}_{-0.38}\right) \times 10^{-4}$} and similar phase of \mbox{$\phi_{\rhoomega} = \left( -39^{+140}_{-69}\right) \degree$} are recovered. The precision of the measured $m_{\pi\pi}$ spectrum, however, is not sufficient to provide a 68\% confidence region on $R$ itself, and we only can determine a range of $R \in [0.4,47] \, \mathrm{GeV}^{-1}$ at 50 \% CL. 
 
 Assuming a fully in-phase production of $\rho$ and $\omega$ in the semileptonic decay by fixing $\phi_\rhoomega = 0$, results in a marginally larger branching fraction of \mbox{$\BFrho = \left( 1.48_{-0.39}^{+0.39} \right) \times 10^{-4}$}. Assuming a phase shift of $\phi_\rhoomega = \pi$, similar to the phase observed in $B^0 \to \overline D^0 \pi^+ \pi^-$ decays~\cite{LHCb:2015klp}, we find \mbox{$\BFrho = \left( 1.35^{+0.38}_{-0.38} \right) \times 10^{-4}$}. Notably the upper uncertainty is reduced, thus theory input on the strong phase difference has the potential to reduce the $\rho$ branching fraction uncertainty. 
 
The branching fraction Eq.~\ref{eq:res_rho_BW} is about 0.2 standard deviations larger than the world average of Eq.~\ref{eq:BFrho} of $\BFrho = \left( 1.35 \pm 0.12 \right) \times 10^{-4}$~\cite{Bernlochner:2021rel}. The two-dimensional allowed 68\% CL region also contains larger branching fractions with values up to $\sim 2.1 \times 10^{-4}$. Combining Eq.~\ref{eq:res_rho_BW} with the form factor predictions of Ref.~\cite{Bharucha:2015bzk} we determine
\begin{align}
 |V_{ub}|_{\rhoomega} & = \left( 3.03^{+0.49}_{-0.44} \right) \times 10^{-3} \, .
\end{align}
This value is about 2\% larger than the world average of Eq.~\ref{eq:Vub_rho}. Taking into account interference effects, we recover an increased upper uncertainty, what reduces the tension from $|V_{ub}|$ from $B \to \pi \ell \bar \nu_\ell$ to $1.3\sigma$. This direct comparison, however, is not well suited to quantify the importance of correctly treating interference effects, as the world average and Eq.~\ref{eq:res_rho_BW} do rely on different assumptions for the subtraction of the $S$-wave semileptonic background. Further, using the measurement of Ref.~\cite{Belle:2021eni} in contrast to an average of many measurements results in a larger overall uncertainty on $|V_{ub}|$.

A better suited comparison to assess the impact of including  the $\rhoomega$ interference effects is to determine the branching  using a simpler resonance model and compare the result with Eq.~\ref{eq:res_rho_BW}. We describe the $\rho$ signal using a relativistic Breit-Wigner, the $S$-wave analogously as before, and neglect the small direct $\omega \to \pi \pi$ contribution. We find
$
\BFrho = \left( 1.48 \pm 0.38 \right) \times 10^{-4} \, .
$
The observed downward shift is $\simeq 0.07 \times 10^{-4}$, corresponding to about 18\% of the quoted uncertainty or about 4.8\% of the central value, respectively. This shift could be used as a proxy to estimate an uncertainty due to the interference for existing measurements.

\begin{table}
\renewcommand{\arraystretch}{1.3}
\caption{ Determined values of all parameters of the fit using the relativistic Breit-Wigner to describe the $\rho$ line shape and the proposed shape of Ref.~\cite{Kang:2013jaa} for the $S$-wave contribution.
}
\begin{tabular}{l|c}
\hline\hline
  Parameter & Value \\
   \hline
   \BFrho & \BFrhoFitValue  \\ 
   $\phi_{\rhoomega}$ & $\PhaseomrhoFitValue$ \\
   $\Delta \mathcal{B}(B \to [\pi^+\pi^-]_S \, \ell \bar \nu_\ell)$ & $\left( 0.29 \pm 0.17 \right) \times 10^{-4}$ \\ 
   \hline 
   $\mathcal{B}(B \to \omega ( \to 2 \pi) \ell \bar \nu_\ell)$ & $\left( 0.017 \pm 0.002 \right) \times 10^{-4}$ \\    
   $M_\rho$ &  $ \left( 0.7603^{+0.0017}_{-0.0015} \right) \, \mathrm{GeV}$  \\
   $\Gamma_\rho$ & $\left( 0.1472^{+0.0020}_{-0.0022} \right) \, \mathrm{GeV}$  \\
   $M_\omega$ &   $ \left( 0.7827 \pm 0001 \right) \, \mathrm{GeV}$ \\
   $\Gamma_\omega$ & $\left( 0.0087 \pm 0.0001  \right)  \, \mathrm{GeV} $  \\   
   $R$ & $5.27^{+0.89}_{-0.70} \, \mathrm{GeV}^{-1} $ \\
   $|\delta|$ & $0.0021^{+0.0004}_{-0.0003} \, \mathrm{GeV} $ \\
   $\arg \delta$ & $0.22 \pm 0.06$ \\
\hline\hline
\end{tabular}
\label{tab:fit_results_rho_BW_SW}
\end{table}

\begin{table}
\renewcommand{\arraystretch}{1.3}
\caption{ The determined $B \to \rho^0 \ell \bar \nu_\ell$ branching fractions and $\phi_{\rhoomega}$ phases using different parameterizations for the $\rho$ line shape are listed. }
\begin{tabular}{l|c|c|c|c}
\hline\hline
 Line shape & Eq.  &  $\BFrho \, [10^{-4}]$  & $\phi_{\rhoomega} \, [\degree]$ & $\chi^2$  \\
   \hline
   Rel. Breit-Wigner & \ref{eq:BW}            &   
  $1.41_{-0.38}^{+0.49}$ & $ -46_{\,-67}^{+155} $  & 2.07 \\
   Dyn. Breit-Wigner & \ref{eq:dynBW}         &  $1.41_{-0.38}^{+0.49}$ & $ -47_{\,-67}^{+156} $ & 2.09 \\
   Gounaris-Sakurai & \ref{eq:GS}             &  $1.42_{-0.38}^{+0.47}$ & $ -44_{\,-67}^{+150} $ & 1.98 \\
   al. Dyn. Breit-Wigner & \ref{eq:dynBWalt} &  $1.41_{-0.38}^{+0.49}$ & $ -47_{\,-67}^{+156} $ & 2.06 \\
   al. Gounaris-Sakurai & \ref{eq:GSalt}     &  $1.42_{-0.38}^{+0.47}$ & $ -44_{\,-67}^{+149} $ & 1.96 \\
\hline\hline
\end{tabular}
\label{tab:fit_results_rho_all_SW}
\end{table}

The impact of different choices to model the $S$-wave contribution, is studied by carrying out fits using either the parameterization of Refs.~\cite{COMPASS:2015gxz,Au:1986vs} or a phase space. With a relativistic Breit-Wigner for the $\rho$ and with Refs.~\cite{COMPASS:2015gxz,Au:1986vs}  we find \mbox{$
 \BFrho  = \left( 1.45_{-0.41}^{+0.61} \right) \times 10^{-4}  $} and $ \phi_{\rhoomega}  = \left( -45_{\,-69}^{+162} \right)\degree$. The upward shift in the $\rho$ branching fraction is caused by the lower number of predicted $S$-wave events below the $\rho$ peak. The recovered phase is in good agreement. Using phase space for the $S$-wave we determine \mbox{$\BFrho = \left( 1.41_{-0.39}^{+0.52} \right) \times 10^{-4}  $} and $
 \phi_{\rhoomega}  = \left( -48_{\,-67}^{+159} \right)\degree$. This branching fraction is nearly identical with Eq.~\ref{eq:res_rho_BW}. The marginal shift in the phase is caused by the shape difference of the total line-shape above the $\rho$ resonance peak. Again, the precise details on what parameterization for the $\rho$ is not important, as long as the same physical pole is enforced. The full details of both sets of fits are summarized in Appendix~\ref{app:A}. 
  
\section{Discussion and Conclusions}

We demonstrated that interference effects and the modeling of $S$-wave contributions can have a sizeable effect on the determination of the $B \to \rho^0 \ell \bar \nu_\ell$ branching fraction. The choice of the precise line shape to describe the $\rho$ resonance, however, has only a negligible impact on the determined branching fractions if mass and width of the $\rho$ resonance are constrained to yield the same physical pole in the $s$ plane. Using the $S$-wave shape of Ref.~\cite{Kang:2013jaa} and a relativistic Breit-Wigner to describe the $\rho$ resonance, we determine with a fit to the measurement of the $m_{\pi\pi}$ spectrum of $B \to \pi \pi \ell \bar \nu_\ell$ of Ref.~\cite{Belle:2020xgu} the $B \to \rho^0 \ell \bar \nu_\ell$ branching fraction, for the first time taking into account the $\rho-\omega$ interference effects in a consistent way to our knowledge. We find 
 \begin{align} \nonumber
 \mathcal{B}(B \to \rho^0 \ell \bar \nu_\ell) & = \BFrhoFitValue  \, , \\ \nonumber
 \phi_{\rhoomega} & = \PhaseomrhoFitValue \, ,
 \end{align}
 and constrain a possible $S$-wave contribution to
 \begin{align}  \nonumber
 \Delta\mathcal{B}(B \to [\pi^+ \pi^-]_S \ell \bar \nu_\ell) & <  \BFpipiFitValue \, ,
 \end{align}
 within $m_{\pi\pi} \in [2 m_\pi, 1.02\, \mathrm{GeV}]$. These values are higher than the world average of Ref.~\cite{Bernlochner:2021rel}, seemingly easing the tension of $|V_{ub}|$ determinations from $B \to \rho \ell \bar \nu_\ell$ with respect to $B \to \pi \ell \bar \nu_\ell$. With this branching fraction and the predictions for the rate of Ref.~\cite{Bharucha:2015bzk} we recover
 \begin{align}
 |V_{ub}|_{\rhoomega} & = \left( 3.03^{+0.49}_{-0.44} \right) \times 10^{-3} \, .
\end{align}
 A comparison using the same data set and assumptions for the $S$-wave $\pi\pi$ contribution reveals that not taking into account interference effects may result in a shift of the order of $\simeq 0.7 \times 10^{-4}$ on the branching fraction. An improved description of the shape of the possible $S$-wave contribution is very important for a reliable determination of the $\rho$ branching fraction. We tested two different models and obsersve that depending on the $S$-wave model the branching fraction may shift up to $\approx 0.03 \times 10^{-4}$ and the phase by $\approx 3 \degree$. The shift in the branching fraction corresponds to about 11\% (14\%) of the obtained upper (lower) uncertainty from the fit. 

 With the arrival of new and enlarged data sets from both Belle~II~\cite{Belle-II:2018jsg} and LHCb~\cite{Kirsebom:2864159}, we must adapt the modelling of the $2 \pi$ mass spectrum for future studies of $B \to \rho \ell \bar \nu_\ell$. Both the $\rhoomega$ interference and the $S$-wave contribution must be taken into account to reduce systematic uncertainties and exploit the expected statistical precision. 
 
 Specifically, if partial branching fractions are measured, which do not integrate the full angular information e.g. due to acceptance effects, additional interference effects also between the $S$-wave and the signal will become important. One possible remedy for existing measurements could be to assign an additional 4\% uncertainty to the measured partial branching fractions, based on the observed shift in analyzing the data set of Ref.~\cite{Belle:2020xgu} with a line shape with and without \rhoomega-interference effects. These studies will complement the golden channel of $B \to \pi \ell \bar \nu_\ell$ to extract the value of $|V_{ub}|$. A more precise understanding of  $B \to \pi\pi \ell \bar \nu_\ell$ decays will also improve future measurements of inclusive semileptonic decays of $B$ mesons, as well as searches for $B \to \mu \bar \nu_\mu$. With more data at hand, the analyses should also exploit angular distributions of the $2 \pi$ system, allowing a clear separation of $S$-, and $P$-wave, as well as other background contributions. 
  
 A more detailed analysis of the $m_{\pi\pi}$ spectrum, which extends the fit to the full measured range,s is left for future work.

\FloatBarrier

\vspace{2ex}

\section*{Acknowledgments} 

The authors want to thank especially Bob Kowalewski for pointing out this important effect and Stephan Paul for detailed feedback on the core content of this manuscript. Further thanks for insightful discussions go to Moritz Bauer and Peter Lewis. We are further indebted to Christoph Schwanda, Dean Robinson, Zoltan Ligeti, Markus Prim, and Svenja Granderath for providing additional input on the manuscript. FB thanks Gilles Tessier and Suzanne Tessier for enlightening and captivating conversations at the lake house about interference. FB is supported by DFG Emmy-Noether Grant No.\ BE~6075/1-1 and BMBF Grant No.\ 05H21PDKBA.

%\nocite{*}

\bibliographystyle{apsrev4-1}
\bibliography{rho-omega-interference}

\clearpage

\onecolumngrid

\begin{appendix}

\section{Alternative Description of the $S$-Wave contribution with phase-space}\label{app:A}

\begin{figure}[h]
\begin{center}
\includegraphics[width=.49\textwidth]{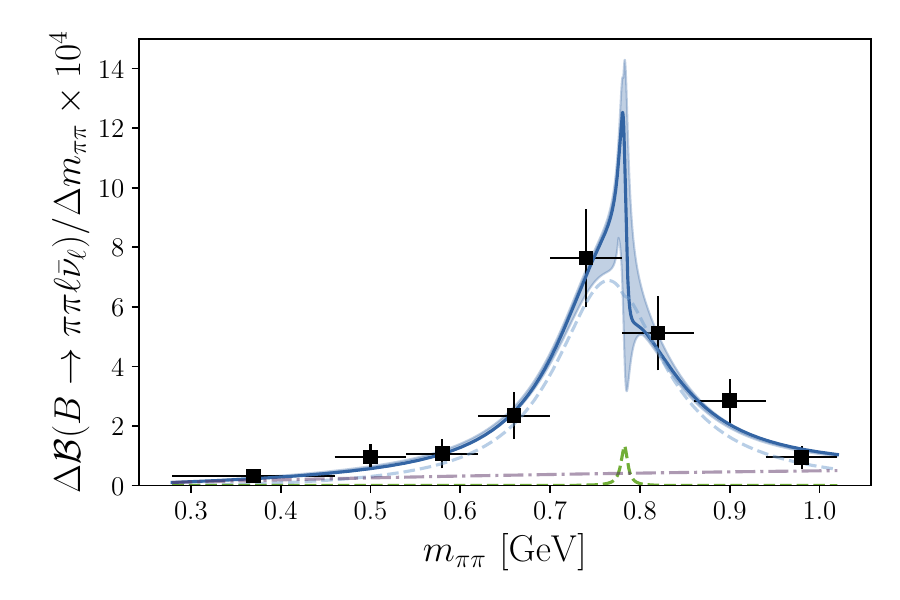}
\includegraphics[width=.49\textwidth]{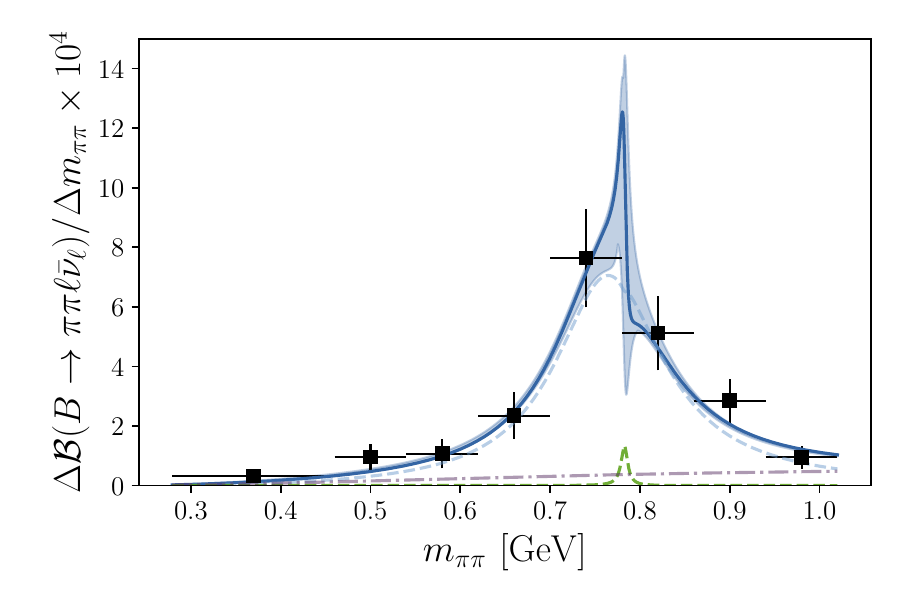}
\caption{ Fit results using the relativistic Breit-Wigner to describe the $\rho$ line shape with using phase space (left) or Ref.~\cite{COMPASS:2015gxz,Au:1986vs} to model the $S$-wave component. The blue line show the full interference and background line shape. The $S$-wave contribution is shown as a dash-dotted purple line. The dashed lines show the prediction without interference for $\rho$ (blue) and $\omega$ (green). }
\label{fig:fit_BW_PS}
\vspace{-4ex}
\end{center}
\end{figure}

\begin{figure}[h]
\begin{center}
\includegraphics[width=.49\textwidth]{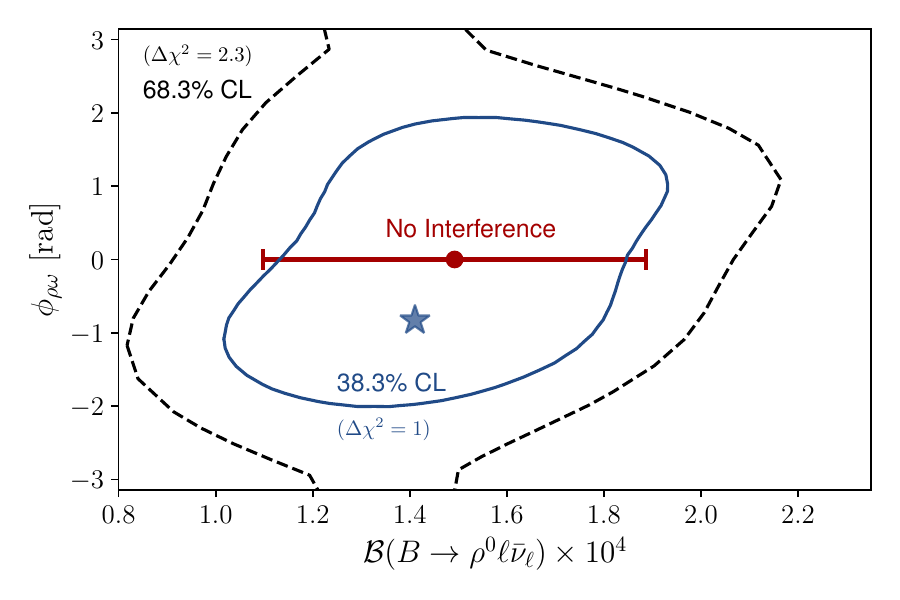}
\includegraphics[width=.49\textwidth]{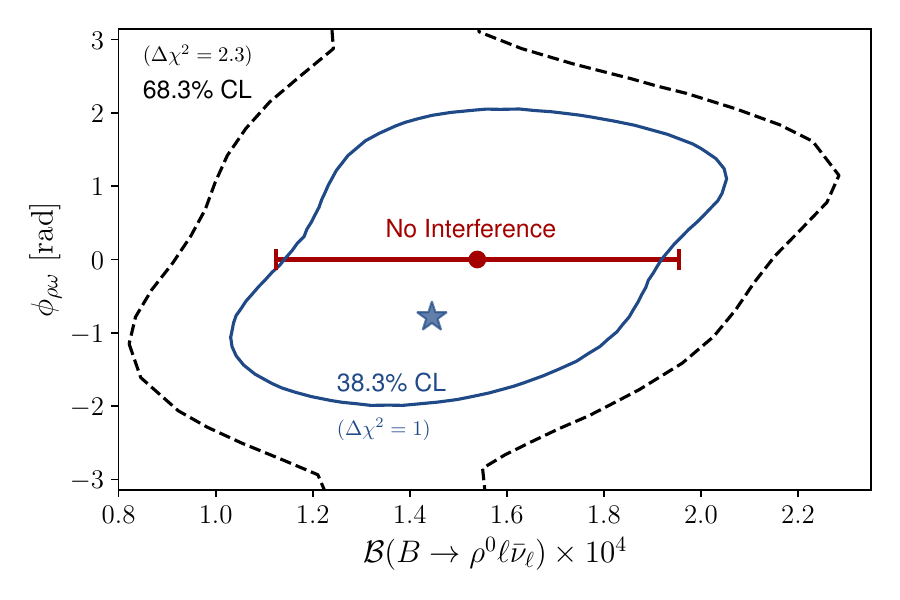}
\caption{ The $\Delta \chi^2 = 1$ (38.3\% CL, blue) and $\Delta \chi^2 =2.3$ (68.3 \% CL, black dashed) contours of \BFrho and $\phi_{\rhoomega}$ are shown for the $S$-wave described using phase space (left) or the model of Ref.~\cite{COMPASS:2015gxz,Au:1986vs} (right). The best fit point ($\Delta \chi^2 = 0$) is indicated with a blue star. The red data point shows the result of a fit neglecting the interference effects between $\rho$ and $\omega$.   }
\label{fig:fit_BW_3}
\vspace{-4ex}
\end{center}
\end{figure}

Figure~\ref{fig:fit_BW_PS} shows the fits to the $m_{\pi\pi}$ spectrum of Ref.~\cite{Belle:2020xgu} using a phase space model or Refs~\cite{COMPASS:2015gxz,Au:1986vs}  for the $S$-wave contribution. The two dimensional $\chi^2$ contours of the determined $\rho$ branching fraction and phase are shown in Figure~\ref{fig:fit_BW_3} for 38.3\% and 68.3\% CL. Tables~\ref{tab:fit_results_rho_BW_PS} summarizes the fitted parameters and Table~\ref{tab:fit_results_rho_all_PS} shows the impact of choosing different parameterizations for the $\rho$ line shape. 

\begin{table}[h]
\renewcommand{\arraystretch}{1.3}
\caption{ Determined values of all parameters of the fit using the relativistic Breit-Wigner to describe the $\rho$ line shape and phase space (left) or Refs.~\cite{COMPASS:2015gxz,Au:1986vs} (right) for the $S$-wave contribution. %The first three listed parameters are completely unconstrained.
}
\vspace{2ex}
\begin{tabular}{l|c}
\hline\hline
  Parameter & Value \\
   \hline
   \BFrho & $\left( 1.41^{+0.52}_{-0.39} \right) \times 10^{-4}$  \\ 
   $\phi_{\rhoomega}$ & $\left(-48^{+160}_{\,-67} \right) \degree$ \\
   $\Delta \mathcal{B}(B \to [\pi^+\pi^-]_S \, \ell \bar \nu_\ell)$ & $\left( 0.25 \pm 0.16 \right) \times 10^{-4}$ \\ 
   \hline 
   $\mathcal{B}(B \to \omega ( \to 2 \pi) \ell \bar \nu_\ell)$ & $\left( 0.017 \pm 0.002 \right) \times 10^{-4}$ \\    
   $M_\rho$ &  $ \left( 0.7603^{+0.0017}_{-0.0015} \right) \, \mathrm{GeV}$  \\
   $\Gamma_\rho$ & $\left( 0.1472^{+0.0020}_{-0.0022} \right) \, \mathrm{GeV}$  \\
   $M_\omega$ &   $ \left( 0.7827 \pm 0001 \right) \, \mathrm{GeV}$ \\
   $\Gamma_\omega$ & $\left( 0.0087 \pm 0.0001  \right)  \, \mathrm{GeV} $  \\   
   $R$ & $5.33^{+0.89}_{-0.71} \, \mathrm{GeV}^{-1} $ \\
   $|\delta|$ & $0.0021^{+0.0003}_{-0.0003} \, \mathrm{GeV} $ \\
   $\arg \delta$ & $0.22 \pm 0.06$ \\
\hline\hline
\end{tabular}
\hspace{2ex}
\begin{tabular}{l|c}
\hline\hline
  Parameter & Value \\
   \hline
   \BFrho & $\left( 1.45^{+0.61}_{-0.41} \right) \times 10^{-4}$  \\ 
   $\phi_{\rhoomega}$ & $\left(-45^{+162}_{\,-69} \right) \degree$ \\
   $\Delta \mathcal{B}(B \to [\pi^+\pi^-]_S \, \ell \bar \nu_\ell)$ & $\left( 0.20^{+0.17}_{-0.23} \right) \times 10^{-4}$ \\ 
   \hline 
   $\mathcal{B}(B \to \omega ( \to 2 \pi) \ell \bar \nu_\ell)$ & $\left( 0.017 \pm 0.002 \right) \times 10^{-4}$ \\    
   $M_\rho$ &  $ \left( 0.7603^{+0.0017}_{-0.0015} \right) \, \mathrm{GeV}$  \\
   $\Gamma_\rho$ & $\left( 0.1472^{+0.0020}_{-0.0022} \right) \, \mathrm{GeV}$  \\
   $M_\omega$ &   $ \left( 0.7827 \pm 0001 \right) \, \mathrm{GeV}$ \\
   $\Gamma_\omega$ & $\left( 0.0087 \pm 0.0001  \right)  \, \mathrm{GeV} $  \\   
   $R$ & $5.33^{+0.89}_{-0.71} \, \mathrm{GeV}^{-1} $ \\
   $|\delta|$ & $0.0021^{+0.0003}_{-0.0003} \, \mathrm{GeV} $ \\
   $\arg \delta$ & $0.22 \pm 0.06$ \\
\hline\hline
\end{tabular}
\label{tab:fit_results_rho_BW_PS}
\end{table}

\begin{table}[h]
\renewcommand{\arraystretch}{1.3}
\caption{ The determined $B \to \rho^0 \ell \bar \nu_\ell$ branching fractions and $\phi_{\rhoomega}$ phases using different parameterizations for the $\rho$ line shape are listed for phase space (top) or Refs.~\cite{COMPASS:2015gxz,Au:1986vs} (bottom) to describe the $S$-wave. }
\vspace{2ex}
\begin{tabular}{l|c|c|c|c}
\hline\hline
 Line shape & Eq.  &  $\BFrho \, [10^{-4}]$  & $\phi_{\rhoomega} \, [\degree]$ & $\chi^2$  \\
   \hline
   Rel. Breit-Wigner & \ref{eq:BW}            &  $1.41_{-0.39}^{+0.52}$ & $-48_{\,-67}^{+159}$ & 2.86 \\
   Dyn. Breit-Wigner & \ref{eq:dynBW}         &  $1.41_{-0.39}^{+0.52}$ & $-48_{\,-67}^{+159}$ & 2.86 \\
   Gounaris-Sakurai & \ref{eq:GS}             &  $1.42_{-0.40}^{+0.51}$ & $-45_{\,-67}^{+153}$ & 2.92 \\
   alt. Dyn. Breit-Wigner & \ref{eq:dynBWalt} &  $1.41_{-0.39}^{+0.52}$ & $-48_{\,-67}^{+159}$ & 2.87 \\
   alt. Gounaris-Sakurai & \ref{eq:GSalt}     &  $1.41_{-0.40}^{+0.40}$ & $-45_{\,-67}^{+153}$ & 2.93 \\
\hline\hline
\end{tabular}

\vspace{3ex}

\begin{tabular}{l|c|c|c|c}
\hline\hline
 Line shape & Eq.  &  $\BFrho \, [10^{-4}]$  & $\phi_{\rhoomega} \, [\degree]$ & $\chi^2$  \\
   \hline
   Rel. Breit-Wigner & \ref{eq:BW}            &  $1.45_{-0.41}^{+0.62}$ & $-45_{\,-69}^{+162}$ & 2.86 \\
   Dyn. Breit-Wigner & \ref{eq:dynBW}         &  $1.44_{-0.41}^{+0.61}$ & $-45_{\,-69}^{+163}$ & 2.86 \\
   Gounaris-Sakurai & \ref{eq:GS}             &  $1.46_{-0.42}^{+0.61}$ & $-42_{\,-69}^{+157}$ & 2.92 \\
   alt. Dyn. Breit-Wigner & \ref{eq:dynBWalt} &  $1.44_{-0.41}^{+0.61}$ & $-45_{\,-69}^{+163}$ & 2.87 \\
   alt. Gounaris-Sakurai & \ref{eq:GSalt}     &  $1.46_{-0.42}^{+0.61}$ & $-42_{\,-69}^{+157}$ & 2.93 \\
\hline\hline
\end{tabular}
\label{tab:fit_results_rho_all_PS}
\end{table}

\section{$R$-dependence on the line shape}\label{app:B}

\begin{figure}[h]
\begin{center}
\includegraphics[width=.49\textwidth]{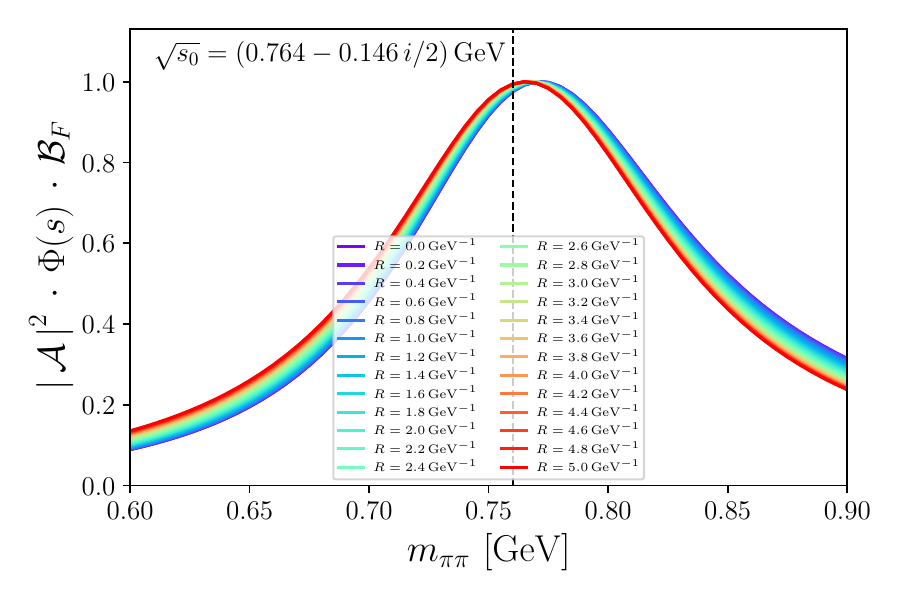}
\includegraphics[width=.49\textwidth]{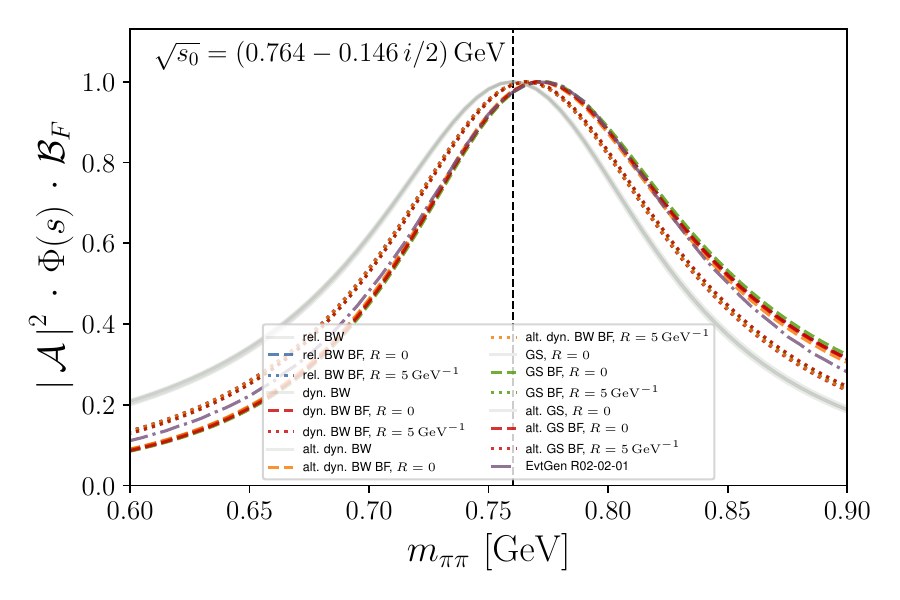}
\caption{ (Left) The influence of changing the scale factor $R$ on the relativistic Breit-Wigner shape is seen, whose size is related to the strong potential.  }
\label{fig:R_variations}
\vspace{-4ex}
\end{center}
\end{figure}

Figure~\ref{fig:R_variations} (left) depicts the impact of different choices of $R$ on the $\rho$ line shape for $R \in [0, 5] \, \mathrm{GeV}^{-1}$, when multiplying the amplitude squared of a relativistic Breit-Wigner with the barrier factor and phase space. Figure~\ref{fig:R_variations} (right) depicts the line shapes for $R = 0$ and $R = 5 \, \mathrm{GeV}^{-1}$ for alternative parameterizations for the $\rho$. All line shapes use $M_0$ and $\Gamma_0$ that reproduce a pole of $\sqrt{s_0} = \left( 0.764 - 0.146 i / 2 \right) \, \mathrm{GeV}$.

\clearpage

\section{ $M_0$ and $\Gamma_0$ values for studied $\rho$ Parameterizations }\label{app:C}

Table~\ref{tab:M0_G0_all_params} lists the values and uncertainties of $M_0$ and $\Gamma_0$ values if the pole of $ \sqrt{s_0} = \left( 763.7^{+1.7}_{-1.5} - 73.2^{+1.0}_{-1.2} i \right) \, \mathrm{MeV}$ is enforced for $R = 5 \, \mathrm{GeV}^{-1}$ (top) and $R = 3 \, \mathrm{GeV}^{-1}$ (bottom). We also list the relativistic Breit-Wigner for comparison, whose parameterization does not depend on $R$.  The recovered values of $M_0$ and $\Gamma_0$ are very weakly correlated with correlation coefficients of $\rho \simeq 0.01\%$. The \texttt{EvtGen} event generator~\cite{EvtGen} implements the dynamic Breit-Wigner Eq.~\ref{eq:dynBW} with a fixed value of $R = 3 \, \mathrm{GeV}^{-1}$, but has default values of $M_0 = 0.77526 \, \mathrm{GeV}$ and $\Gamma_0 = 0.1474 \, \mathrm{GeV}$ which do not reproduce the pole of $ \sqrt{s_0} = \left( 763.7^{+1.7}_{-1.5} - 73.2^{+1.0}_{-1.2} i \right) \, \mathrm{MeV}$.

\begin{table}[t!]
\renewcommand{\arraystretch}{1.3}
\caption{  The recovered $M_0$ and $\Gamma_0$ values for a fit to the pole of $ \sqrt{s_0} = \left( 763.7^{+1.7}_{-1.5} - 73.2^{+1.0}_{-1.2} i \right) \, \mathrm{MeV}$  \cite{Garcia-Martin:2011nna} with $R = 5 \, \mathrm{GeV}^{-1}$ (top) and $R = 3 \, \mathrm{GeV}^{-1}$ (bottom).
}
\vspace{2ex}
\begin{tabular}{l|c|c|c}
\hline\hline
 Line shapes with $R = 5\, \mathrm{GeV}^{-1}$ & Eq.  &  $M_0$ [GeV]  & $\Gamma_0$ [GeV] \\
   \hline
   Rel. Breit-Wigner & \ref{eq:BW}            &  $0.7602 \pm 0.0017$ & $0.1471 \pm 0.0022$  \\
   Dyn. Breit-Wigner & \ref{eq:dynBW}         & $0.7651 \pm 0.0017$ & $0.1452 \pm 0.0020$  \\
   Gounaris-Sakurai & \ref{eq:GS}             & $0.7657 \pm 0.0017$ &  $0.1453 \pm 0.0020$ \\
   alt. Dyn. Breit-Wigner & \ref{eq:dynBWalt} & $0.7721 \pm 0.0015$ &$0.1479 \pm 0.0022$  \\
   alt. Gounaris-Sakurai & \ref{eq:GSalt}     & $0.7728 \pm 0.0015$ & $ 0.1482 \pm 0.0021$  \\
\hline\hline
\end{tabular}
\vspace{2ex}
\\
\begin{tabular}{l|c|c|c}
\hline\hline
 Line shapes with $R = 3\, \mathrm{GeV}^{-1}$  & Eq.  &  $M_0$ [GeV]  & $\Gamma_0$ [GeV] \\
   \hline
   Rel. Breit-Wigner & \ref{eq:BW}            &  $0.7602 \pm 0.0017$ & $0.1471 \pm 0.0022$  \\
   Dyn. Breit-Wigner & \ref{eq:dynBW}        & $0.7688 \pm 0.0017$ & $0.1455 \pm 0.0022$  \\
   Gounaris-Sakurai & \ref{eq:GS}             & $0.7694 \pm 0.0017$ &  $0.1457 \pm 0.0020$ \\
   alt. Dyn. Breit-Wigner & \ref{eq:dynBWalt} & $0.7759 \pm 0.0015$ &$0.1496 \pm 0.0023$  \\
   alt. Gounaris-Sakurai & \ref{eq:GSalt}     & $0.7765 \pm 0.0016$ & $ 0.1502 \pm 0.0021$  \\
\hline\hline
\end{tabular}
\label{tab:M0_G0_all_params}
\end{table}

\section{ Calculation of Branching Fractions }\label{app:D}

We choose $\mathcal{B}_\rho$ and $|B|$ such that the interference amplitude Eq.~\ref{eq:A_r_w} reproduces
\begin{align}
 \mathcal{B}(B \to \omega (\to \pi \pi) \ell \bar \nu_\ell) & = \mathcal{B}_\rho 
  \int\limits_{2 m_\pi}^{\infty}  \, \mathrm{d} m_{\pi\pi} \left| \frac{ \mathcal{A}_{\rho}(m_{\pi\pi}^2)  \mathcal{A}_{\omega}(m_{\pi\pi}^2) \, \Delta \, |B| e^{i \phi_{\rhoomega}}}{1 - \Delta^2 \mathcal{A}_{\rho}(m_{\pi\pi}^2) \mathcal{A}_{\omega}(m_{\pi\pi}^2)}   \right|^2 \cdot  \Phi(m_{\pi\pi}^2) \cdot \mathcal B_F(m_{\pi\pi}^2)  \, , \\
 \BFrho & = \mathcal{B}_\rho 
  \int\limits_{2 m_\pi}^{\infty}  \, \mathrm{d} m_{\pi\pi} \left| \frac{ \mathcal{A}_{\rho}(m_{\pi\pi}^2)}{1 - \Delta^2 \mathcal{A}_{\rho}(m_{\pi\pi}^2) \mathcal{A}_{\omega}(m_{\pi\pi}^2)}   \right|^2 \cdot  \Phi(m_{\pi\pi}^2) \cdot \mathcal B_F(m_{\pi\pi}^2)  \, .
\end{align}

We choose for the $S$-wave $\mathcal B_s$ such that 
\begin{align}
\Delta \mathcal{B}(B \to [\pi\pi]_S \ell \bar \nu_\ell) & = \mathcal{B}_S 
  \int\limits_{2 m_\pi}^{1.02 \, \mathrm{GeV}}  \, \mathrm{d} m_{\pi\pi} \, \bigg| \mathcal{A}_{S-\mathrm{wave}}(m_{\pi\pi}^2) \bigg|^2 \cdot  \Phi(m_{\pi\pi}^2)  \, .
\end{align}

%  \mathcal{A}_{\rhoomega} = \mathcal{A}_{\rho} \left( \frac{1 + \mathcal{A}_{\omega} \, \Delta \, |B| e^{i \phi_{\rhoomega}}}{1 - \Delta^2 \mathcal{A}_{\rho} \mathcal{A}_{\omega}}  \right) \, ,

% \right .\\[-1em]
%$$        &\left. \quad\,\, + \quad \mathcal{B}_{S} \cdot \left|\mathcal A_{S-\mathrm{wave}}\right|^2 \Phi(s) \,\, \right) \, .

\end{appendix}

\end{document}